\documentclass[a4paper,fleqn,usenatbib]{mnras}

\usepackage[T1]{fontenc}
\usepackage{ae,aecompl}

\usepackage{graphicx}	
\usepackage{amsmath}	
\usepackage{amssymb}	

\title[On the Alfv\'enic Slow Solar Wind]{On slow solar wind with high Alfv\'enicity: from composition and microphysics to spectral properties}

\author[R. D'Amicis et al.]{
Raffaella D'Amicis,$^{1,2}$\thanks{E-mail: raffaella.damicis@inaf.it}
Lorenzo Matteini,$^{3}$
and Roberto Bruno$^{2}$
\\
$^{1}$National Institute for Astrophysics (INAF), Institute for Space Astrophysics and Planetology (IAPS), Via del Fosso del Cavaliere 100,\\
 00133 Rome, Italy\\
$^{2}$Serco S.p.A, Via Sciadonna 24, 00044, Frascati, Italy\\
$^{3}$LESIA, Observatoire de Paris, Universit\'e PSL, CNRS, Sorbonne Universit\'e, Univ. Paris Diderot, Sorbonne Paris Cit\'e, 5 place Jules \\
 Janssen, 92195 Meudon, France
}

\date{Accepted 04 December 2018. Received 29 November; in original form 06 September 2018}

\pubyear{2018}

\begin{document}
\label{firstpage}
\pagerange{\pageref{firstpage}--\pageref{lastpage}}
\maketitle

\begin{abstract}
Alfv\'enic fluctuations are very common features in the solar wind and are found especially within the main portion of fast wind streams while the slow wind usually is less Alfv\'enic and more variable. In general, fast and slow wind show many differences which span from the large scale structure to small scale phenomena including also a different turbulent behaviour. Recent studies, however, have shown that even slow wind can be sometimes highly Alfv\'enic with fluctuations as large as those of the fast wind. The present study is devoted to present many facets of this Alfv\'enic slow solar wind including for example the study of the source regions and their connection to coronal structures, large-scale properties and micro-scale phenomena and also impact on the spectral features. This study will be conducted performing a comparative analysis with the typical slow wind and with the fast wind. 
It has been found that the fast wind and the Alfv\'enic slow wind share common characteristics, probably attributable to their similar solar origin, i.e. coronal-hole solar wind. Given these similarities, it is suggested that in the Alfv\'enic slow wind a major role is played by the super-radial expansion responsible for the lower velocity. Relevant implications of these new findings for the upcoming Solar Orbiter and Solar Probe Plus missions, and more in general for turbulence measurements close to the Sun, will be discussed.
\end{abstract}




\section{Introduction}

The solar wind comes mainly into two distinct flavors: fast and slow speed solar wind \citep[e.g.][]{bd}, showing differences which extend well beyond their speeds.
The fast solar wind, typically associated with speeds exceeding 600 km/s, originates primarily from coronal holes. It is also characterized by ion temperatures that far exceed electron temperatures in the inner corona, at least out to 10 $R_S$ from the Sun. On the other hand, the near-ecliptic slow solar wind has characteristics that are distinct from the fast wind: its speed is typically $<$ 500 km s$^{-1}$, and the ion temperature tends to be lower than the electron temperature. The properties of the slow solar wind are far more dynamic and variable than those of the fast solar wind. The slow solar wind is generally found in the vicinity of the heliospheric current sheet \citep{smith1978} emanating from streamers at the Sun, especially at the time of solar minimum. However, the sources for this slow solar wind have not been clearly established and are still debated \citep[e.g.][]{abbo,antiochos,antonucci,ofman,noci,wangshiley}. Near solar maximum, the slow solar wind may not even be spatially limited to the heliospheric current sheet. 

The composition in the two wind types is clearly
different: in slow speed wind, elements having low first
ionization potential (FIP $\leq$ 10 eV - i.e. low-FIP elements)
are enhanced by a factor of 3 - 4 relative to the photosphere being more representative of closed magnetic structures in the corona,
while in the fast speed their abundances are nearly photospheric
\citep{geiss1995,vonsteiger2000,zurbuchen1999}. Additionally, in the slow solar wind the
freeze-in temperature from carbon charge-states is of
the order of 1.4 -- 1.6 $\times 10^6$ K and in the fast wind it is
8 $\times 10^5$ K. 

Fast and slow wind also differs in terms of the local microphysics.
This concerns the thermodynamic state of all main species that compose the plasma (electrons, protons and alpha particles), as well as their relative streaming.
Different electron populations (core, halo, strahl) have different properties in the two regimes, and display different temperature anisotropies with respect to the magnetic field direction \citep[e.g.,][]{stverak2008}, especially due to the different collisionality of the lighter and hotter fast wind, and the denser and cooler slow one.
Such differences are even more apparent for ions: alpha particles are typically hotter than protons in the fast wind, where they also stream faster than the proton bulk velocity by approximatively the Alfv\'en speed \citep{marsch1982a, neugebauer1996}, unlike typical slow wind where the two populations often display same temperature and no differential speed \citep{kasper2008, maruca2012}.
In particular, ions show strong temperature anisotropies with respect to the local magnetic field direction \citep{marsch1982b, kasper2002}. While protons in fast streams are characterised by large anisotropies with $T_\perp > T_\parallel$ and the presence of a secondary beam population, typical slow wind plasma show $T_\perp \sim T_\parallel$ and no relevant beams \citep{hellinger, matteini2013, marsch1982b}.

Solar wind plasma and magnetic field power density spectra extend over several frequency decades, reflecting the large extension of typical solar wind timescales, and are characterized by different frequency regimes.
Solar wind fluctuations, either generated within fast or slow wind streams, show a typical Kolmogorov-like power law in the inertial range. 
Fast wind shows a $k^{-1}$ scaling at low frequencies which is typical of the large-scale energy
containing eddies. A frequency break separates the $k^{-5/3}$ from the $k^{-1}$ scaling. This break corresponds to the correlation length and moves to ever larger scales as the wind expands in accordance with an increase of the correlation length when moving away from the Sun \citep[see][for related literature]{brunocarbone2013}. This spectral break has been interpreted as evidence that non-linear processes are at work and govern the evolution of solar wind fluctuations \citep{tumarsch1992}. The origin of this $k^{-1}$ scaling is still debated \citep[e.g.][submitted]{mg1986,dmitruk2007,verdini2012,tsuru2018,matteini}. 
Conversely, slow wind shows a $k^{-5/3}$ scaling extending over several frequency decades and, up to recent findings \citep{rb2017}, there was no evidence of the existence of the $k^{-1}$ regime. Moreover, no radial evolution is observed. This is an indication that slow wind turbulence is already fully developed close to the Sun \citep{marsch1990}. 

This behaviour is widely considered to result from the energy cascade process caused by the nonlinear interaction between the inward and outward propagating Alfv\'en waves although the nature and existence of the inward modes are still matter of debate. 
Some evidence of their existence has been provided by \cite{roberts,roberts2,bavassano1989,he2015} although \cite{tsuru2018} suggest caution when intrepreting the results derived from Els\"asser variable analysis. There are also many clues which would suggest that these fluctuations, in some cases, might have a non-Alfv\'enic nature. Several studies on this topic \citep[see the reviews by][]{tumarsch1995,brunocarbone2013} in the low frequency range suggested that non propagating structures probably advected by the wind or locally generated could well act as inward propagating modes. Unfortunately, one-single point measurements do not allow to separate temporal from spatial phenomena and this ambiguity remains unsolved. 

Even if it is out of the scope of this paper to review the theory on the origin of solar wind turbulence, it is worth mentioning other mechanisms causing solar wind turbulence as suggested for example by \cite{tsuru2018}. According to these authors the turbulence spectrum observed by Ulysses in the fast wind could have been generated by a series of stepened Alfv\'en waves, which, similarly to what happens for stepened magnetosonic waves at comets, are able to generate the low and high frequency tail observed in the spectra.

Alfv\'enic correlations are ubiquitous in the solar wind and these
correlations are much stronger and have larger fluctuations amplitude, at lower and lower frequencies as the heliocentric distances become shorter and shorter. 
Actually, \cite{bd} and \cite{bs} showed that, in about 25 days of data from
Mariner 5, out of the 160 days of the whole mission, a strong correlation exists between
velocity and magnetic field fluctuations interpreted as outward propagating (with
respect to the Sun) Alfv\'enic fluctuations. Moreover, in the regions where this correlation is verified to a high degree,
the magnetic field magnitude and number density is almost constant.
Alfv\'enic correlations
are much stronger within the main portion of fast streams, while they
are weak in intervals of slow wind \citep{bd,bs}.
Several authors have found results supporting the idea of an older turbulence in the slow wind and an Alfv\'enic younger turbulence in the fast wind \citep[e.g.][]{tumarsch1995}.
The degree of Alfv\'enic correlations, however, unavoidably fades away with increasing heliocentric 
distance. It must be reported however that there are cases when the absence of strong velocity
shears and compressive phenomena favour a high Alfv\'enic correlation up to very large distances
from the Sun \citep{roberts}.

The turbulent dynamics transfers energy from larger to smaller scales to be eventually dissipated at kinetic
scales. As a matter of fact, around the proton scales, another spectral break is found beyond which the spectrum generally
steepens even if fast and slow wind are characterized by different slopes in the kinetic regime. This part of the spectrum is commonly called the
"dissipation range", in analogy to hydrodynamics, although the
nature of this high-frequency part of the interplanetary fluctuations
is still largely debated \citep[e.g.][]{alexandrova2013,brunocarbone2013}.

Along with this standard classification in fast and slow solar wind, another type of solar wind was found by \cite{damicis2011}, followed by a first attempt to characterize it by \cite{damicis2015}. Actually these authors performed a statistical study to characterize the state of solar wind turbulence at different phases of the solar cycle 23. Using the invariants for the ideal equations of motion, in terms of the normalized cross-helicity, $\sigma_C$ and residual energy $\sigma_R$, it was found that, while at solar minimum the solar wind structure is bimodal with fast wind being more Alfv\'enic (higher $\sigma_C$ values) than slow wind, at solar maximum the expected predominance of slow wind was associated to a population unexpectedly characterized by a high degree of Alfv\'enicity in contrast with \cite{bd} and \cite{bs}. This Alfv\'enic slow solar wind found during maximum of solar cycle 23 will be the focus of the present study. 

The existence of a slow wind showing a high degree of Alfv\'enicity was pointed out first by \cite{marsch1981}. They highlighted the presence of an Afv\'enic slow solar wind during the perihelion passage of Helios 2 from April to May in 1978, in the ascending phase of the solar cycle. During that time interval, no pronounced recurrent high speed streams were observed, such as typically occurred during the period of solar activity minimum in 1974-1976. On the other hand, the low speed wind in 1978 showed highly irregular speed and temperature profiles with numerous short-lived, marked velocity fluctuations (as have usually been observed in the body of fast streams) indicating that this slow wind was basically different
from the slow solar wind at solar activity minimum. Moreover proton velocity distribution functions of this kind of slow wind were found to have 
signatures similar to typical high-speed proton distributions detected during solar activity minimum \citep{feldman1973, feldman1974, marsch1982b}. 

\cite{marsch1981}, however, did not perform a comprehensive characterization of this slow wind. Their findings were derived from measurements quite close to the Sun (around 0.28 AU). On the contrary, the results by \cite{damicis2011} and \cite{damicis2015} were found at 1 AU where we expect a degradation of the v-b correlation. These arguments motivated a further characterization of this kind of slow wind which will be the aim of the present paper.

In particular, section 2 will be dedicated to the description of a case study given as an example and chosen as representative  with particular emphasis on some basics plasma parameters and discussion on the origin of the different solar wind regimes. In section 3 a further characterization of this time interval will follow using derived quantities such as magnetic field compressibility, collisional age, thermal and Alfv\'en speed (and plasma $\beta$). In sections 4 and 5 some differences in the microphysics will be highlighted studying in particular the relationship between $\beta_{p \parallel}$ and $T_\perp/T_\parallel$ and between $\Theta_{BR}$ and $V_{SW}$. In Section 6 a study of the spectral features of magnetic field fluctuations will be performed while section 7 will focus in particular on a study of the location of the spectral break between fluid and kinetic regimes. Summary and conclusions are reported in section 8.

\section{Data selection and main characteristics}
The present analysis is based on data detected during maximum of solar cycle 23 spanning the time window from DoY 1/2000 to 181/2002. We use 24 s moments of the proton velocity distribution function sampled by the experiment Three-Dimensional Plasma and Energetic Particle Investigation (3DP) \citep{lin1995} from the WIND s/c which also includes magnetic field measurements calculated from 3-sec Magnetic Field Investigation (MFI) experiment \citep{lepping1995} on board WIND s/c and averaged over plasma measurements.

\begin{figure*}
	\centerline{\includegraphics[width=400pt]{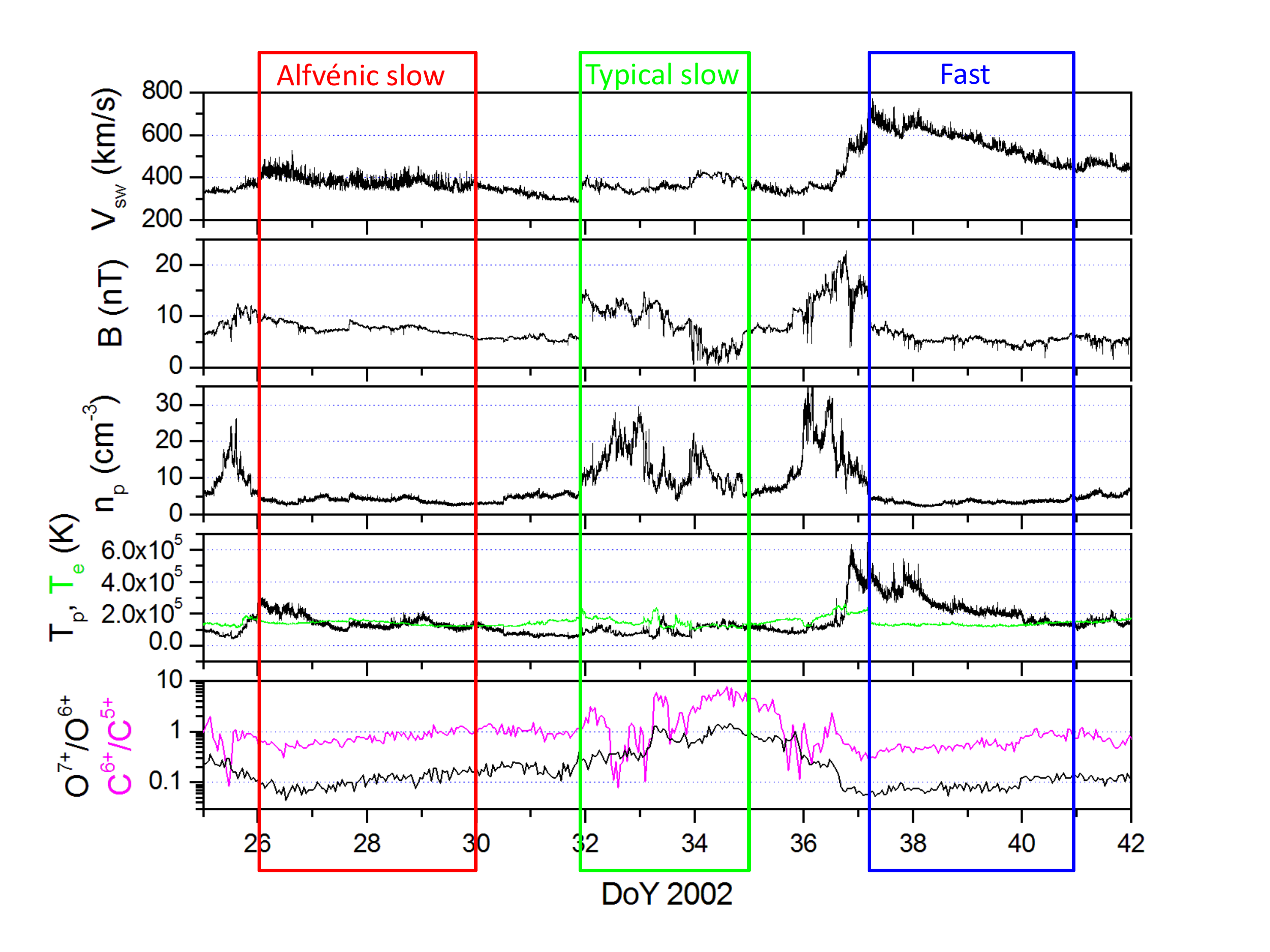}}
	\caption{From top to bottom: solar wind bulk speed, V$_{sw}$ (in km s$^{-1}$), magnetic field magnitude, B (in nT), number density, n$_p$ (in cm$^{-3}$) and proton (black line) and electron (green line) temperature, T$_p$ and T$_e$ (in K), obtained from the 3DP intrument on board WIND s/c. Lower panel: oxygen ratio (O$^{7+}$/O$^{6+}$) (black line) and carbon ratio (C$^{6+}$/C$^{5+}$) (purple line) as derived from 1 hr measurements of SWICS instrument on board ACE s/c. Typical slow, fast and Alfv\'enic slow intervals are indicated with green, blue and red boxes, respectively.}
	\label{figure1}
\end{figure*}

Figure~\ref{figure1} shows a case study corresponding to a time interval ranging from DoY 25 to DoY 42 of 2002. 
This time window was chosen as representative being a good example of three distinct solar wind regimes: a fast wind stream and two slow wind periods. The two slow wind intervals are found to be substantially different from one another from many points of view as we will see in this paper. 

From the first solar wind observations \citep[e.g.][]{schwenn1990}, it was found that basic differences exist between fast and slow solar wind. As mentioned in the introduction, the high-speed wind is characterized by a high proton temperature, a low density,
and a low mass flux, while the low-speed wind is cooler, denser, and has a larger mass flux. Other differences, such as e.g. composition, anisotropies in proton and electron temperatures, have also been observed between slow and fast solar winds. Another relevant difference is found in the Alfv\'enic content with fast wind more Alfv\'enic than slow wind \citep{brunocarbone2013}. 

\cite{damicis2015} and \cite{damicis2016} identified the two slow wind periods as typical (or standard) and Alfv\'enic slow, respectively, indicated also as 1st type and 2nd type, respectively characterized by different features. 

Figure~\ref{figure1} shows, for the selected time interval, from top to bottom: solar wind bulk speed, V$_{sw}$ (in km s$^{-1}$), magnetic field magnitude, B (in nT), number density, n$_p$ (in cm$^{-3}$), proton and electron temperatures, T$_p$ and T$_e$ (in K). 

The Alfv\'enic slow wind has similar bulk speed values respect to the typical slow wind but with larger fluctuations. This is due to the presence of Alfv\'enic fluctuations. Actually, Alfv\'enic periods are usually well detectable also by looking at velocity fluctuations which are enhanced respect to periods of low Alfv\'enic correlations. 

The two Alfv\'enic periods (fast and slow) are characterized by low compression in both number density and magnetic field magnitude as expected. Conversely, the typical slow wind is more variable and high compressions are the main feature of this intervals.

Another important plasma parameter is the proton temperature, T$_p$. Fast wind has higher proton temperature then the slow wind. When looking at the Alfv\'enic slow wind, it is important to notice that the temperature value is midway between that of the typical slow wind (lowest) and that of the fast wind (highest). These two characteristics could be justified supposing a larger expansion rate with respect to the typical Alfv\'enic fast wind.
Moreover, we show also electron temperarure, T$_e$, for reference showing that, as introduced in section 1, in fast wind, ion temperatures exceed electron temperatures while in slow wind the ion temperature are lower than the electron temperature. In the Alfv\'enic slow wind we observe basically similarities with the fast wind even if the trend is not clear along the whole time interval.

The bottom panel of Figure~\ref{figure1} shows oxygen ratio (O$^{7+}$/O$^{6+}$) (black line) and carbon ratio (C$^{6+}$/C$^{5+}$) (purple line) as derived from 1 hr measurements of the Solar Wind Ionic Composition Spectrometer (SWICS) \citep{gloeckler} instrument on board ACE s/c.

Composition observations are particularly useful for identifying the source of solar wind plasma streams. Several studies \citep[e.g.][and reference therein]{geiss1995} suggest a different solar origin for the different solar wind regimes, which has a role in their successive evolution. As a matter of fact, plasma composition is determined close to the Sun by plasma processes occurring in the upper chromosphere near the transition region, and by the temperature history of the plasma between the transition region and 3 solar radii. These plasma processes near the source differ remarkably between the fast and the slow solar wind.

During the expansion process from the coronal source, the ionic charge state adapts to the environment until the recombination and ionization time-scale of a certain ionic charge state becomes large compared to the expansion time-scale. At this point, the respective charge state freezes in as well as the information on the solar wind source region and the expansion properties close to the Sun \citep[see e.g.][]{burgi}. For a given speed, temperature and density profile each ionic charge state has its own specific freeze-in point
which may vary quite considerably from one ion species to another. It turns out that O$^{7+}$/O$^{6+}$, as well as C$^{6+}$/C$^{5+}$ freeze in rather close to the solar wind source region, and therefore show the most variability. Some authors \citep{geiss1995, vonsteiger1997, vonsteiger2000} have pointed out that the charge state composition clearly distinguishes coronal hole associated solar wind from streamer-associated slow solar wind. Furthermore, clear variations within low-speed solar wind also separate different sources of low-speed solar wind \citep{zurbuchen2000}. 

\begin{figure}
	\centerline{\includegraphics[width=250pt]{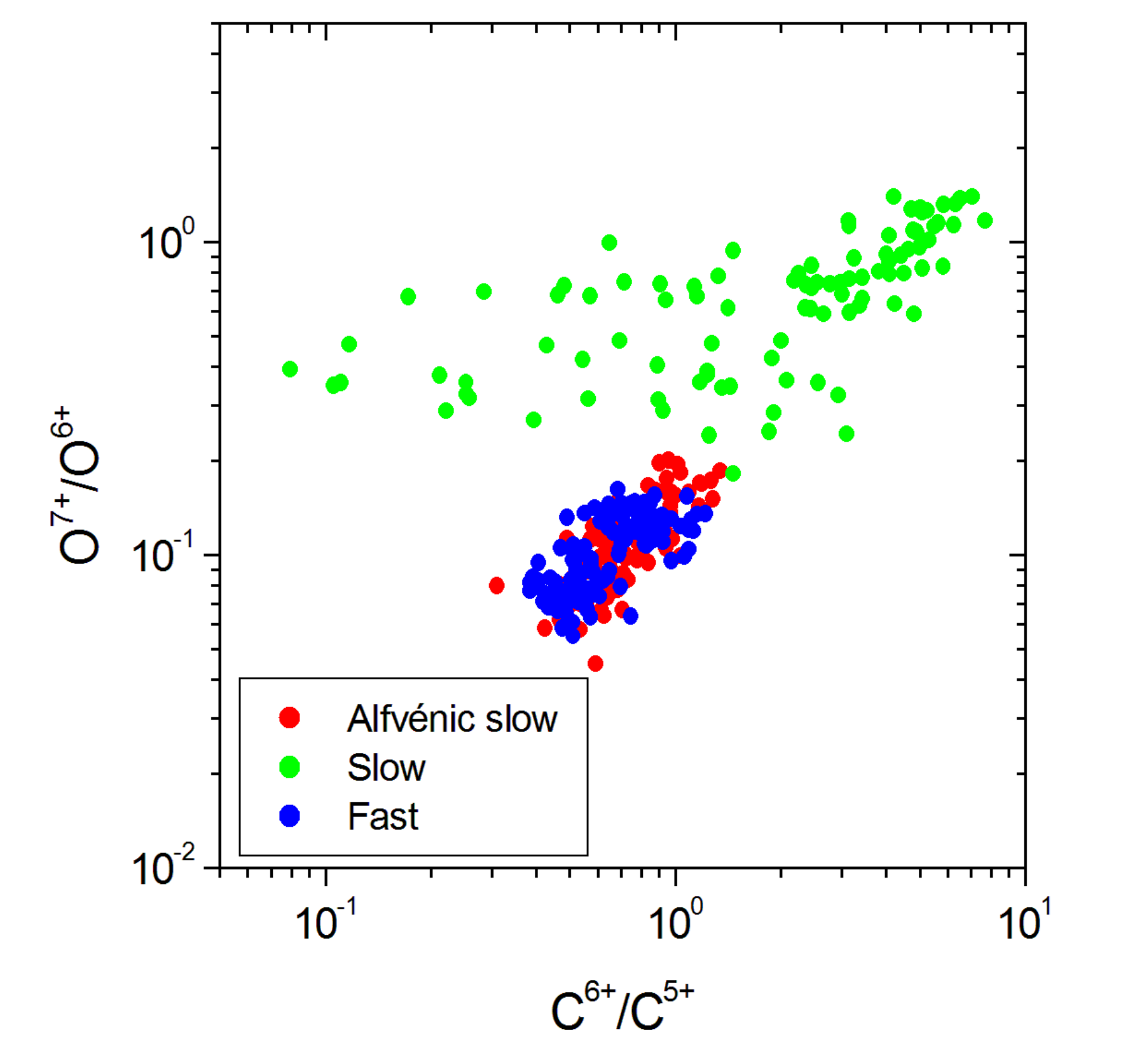}} 
	\caption{Scatter plot of oxygen ratio (O$^{7+}$/O$^{6+}$) and carbon ratio (C$^{6+}$/C$^{5+}$) for typical slow (green dots), fast (blue dots) and Alfv\'enic slow (red dots) solar wind.}
	\label{figure2}
\end{figure}

Results from the present study show unambiguously that the Alfv\'enic time intervals are characterized by lower O$^{7+}$/O$^{6+}$ and C$^{6+}$/C$^{5+}$. \cite{vonsteiger2008}, using Ulysses data, found a well defined clusterization of these two quantities depending on the solar wind regime observed, with typical slow wind characterized by higher O$^{7+}$/O$^{6+}$ and C$^{6+}$/C$^{5+}$ respect to the fast wind. Our study confirms basically previous findings as shown in Figure~\ref{figure2}. Interesting enough is the superposition we observe of the two Alfv\'enic winds, either fast or slow, clearly demonstrating a similar solar origin of the two plasma flows.

The different values of the O$^{7+}$/O$^{6+}$ and C$^{6+}$/C$^{5+}$ ratios indicate plasma coming from different source regions that can be identified by mapping back the measurements on a synoptic map as in \cite{damicis2015} and \cite{damicis2016}. They pointed out that the different solar wind regimes come from different source regions, finding in particular that fast and Alfv\'enic slow streams come from the meridional extensions of the polar coronal holes characterized by open field line regions while the typical slow wind comes from a source region, limited in extension and characterized by a more complex field line topology. These findings all support results by \cite{antonucci} who, using UVCS observations, found evidence for the existence of two kinds of slow solar wind typically originating from different source regions, coronal streamers, and coronal holes' boundaries or from small coronal holes, respectively. It is interesting to notice that the Alfv\'enic slow wind shows plasma features quite similar to those pertaining to the fast wind.  

It is worth stressing that this solar rotation is not peculiar and does not represent an isolated case since \cite{damicis2015} examined all the synoptic maps from Carrington rotations 1958 -- 1991 (during maximum of solar cycle 23) and found very stable magnetic configurations during successive solar rotations. A previous study by \cite{platten} found a remarkable number of localized coronal holes at all latitudes
during the maximum of cycle 23, supporting the findings by \cite{damicis2015} on a statistical basis. This is also in accordance with \cite{wang1994} who identified the origin of the slow wind at solar maximum with small, isolated holes scattered over a wide range of latitudes.

\section{Additional properties} 

Within this context, the idea is to analyse other solar wind features and extend previous studies on the differences between fast and (typical) slow solar wind to the Alfv\'enic slow wind. 

\begin{figure*}
	\centerline{\includegraphics[width=400pt]{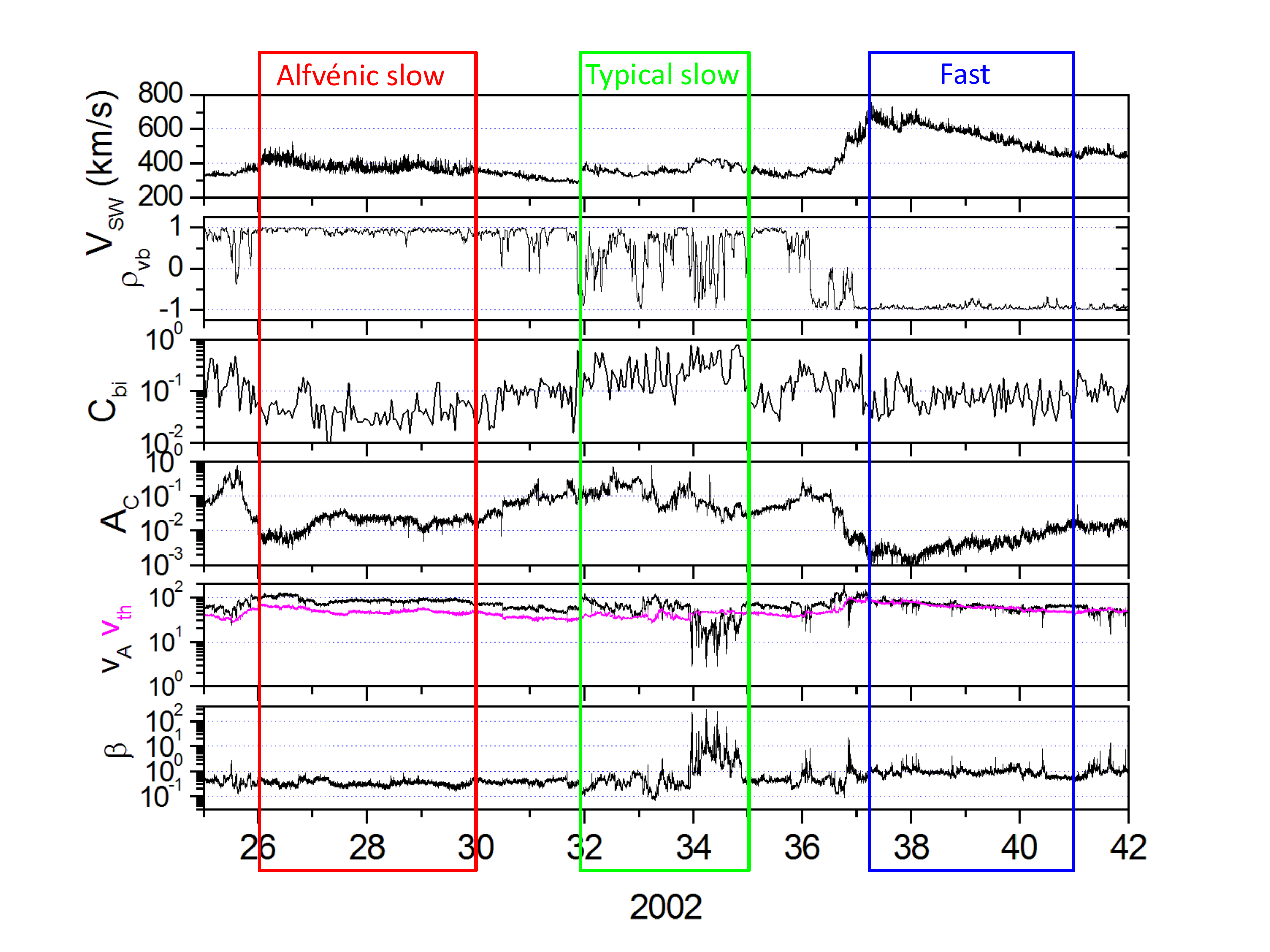}}
	\caption{From top to bottom: solar wind speed profile $V_{sw}$, correlation coefficient between the z components of magnetic field b and velocity v, $R_{vb}$, magnetic field compressibility, $C_{b_i}$, collisional age, $A_C$, thermal speed, $V_{th}$ (km/s) (purple line) and Alfv\'en speed, $V_{A}$ (km s$^{-1}$) (black line), plasma $\beta$. Typical slow, fast and Alfv\'enic slow intervals are indicated with green, blue and red boxes, respectively.}
	\label{figure3}
\end{figure*}

Figure~\ref{figure3} shows from top to bottom: solar wind speed profile, $V_{sw}$ (inserted again for clarity), the correlation coefficient between the z components of magnetic field (b$_z$) and velocity (v$_z$), $R_{vb}$, magnetic field compressibility $C_{b_i}$, the collisional age, $A_C$, the thermal speed, $V_{th}$ (km/s) and the Alfv\'en speed, $V_{A}$ (km s$^{-1}$), and also the plasma $\beta$. These quantities were chosen since they allow a better characterization of the differences between the different solar wind regimes under study.

$R_{vb}$ was computed at 1 hr scale using a sliding window since solar wind fluctuations show a strong Alfv\'enic character at this scale \citep{tumarsch1995}. $R_{vb}$ quantifies the degree of Alfv\'enicity. 
The long lasting high $R_{vb}$ values (absolute value close to 0.9) are found in correspondence of the main portion of the fast wind for DoY 37-42 and for DoY 26-30 corresponding to the Alfv\'enic slow wind. There are also other small and limited intervals of the time series showing very good Alfv\'enic correlations, even if less pure (smaller $R_{vb}$) than the ones characterizing the previous time periods. The typical slow wind on the contrary is characterized by a low degree of Alfv\'enicity and usually the correlation coefficient shows abrupt oscillations from negative to positive values as for example during DoY 32-35.

\cite{brubav1991} suggested that compressible phenomena play an important role in determining the Alfv\'enicity of solar wind fluctuations. 
They studied separately the behaviour of inward and outward modes and found that the depletion of v-b correlations is related to the presence of compressive fluctuations: fluctuations in field intensity produce a depletion of the outward modes while fluctuations in density reinforce the inward mode. In the same study, the authors recommended also caution in defining inward modes simply as inwards Alfv\'en waves since they might be representative of plasmas structures convected by the solar wind rather than propagating waves. The nature of these compressible phenomena was investigated for example by \cite{marschtu1993, tumarsch1994} who found that compressive fluctuations are a complex superposition of magnetoacoustic fluctuations and pressure balance structures \citep{yao2011} whose origin might be local, due to stream dynamical interaction, or of coronal origin related to the flow tube structure.  However, there is little evidence of the presence of magnetosonic waves in the solar wind and in fact, \cite{tsuru2018} found that magnetic decreases (MDs), produced by the steepening of the kinetic Alfv\'en waves, rather than by magnetosonic waves, dominate the compressibility of the interplanetary medium. The reader interested in a more detailed description of this topic can refer to the bibliographic references above and the review by \cite{brunocarbone2013} and references therein.

To better characterize the different solar wind regimes, we computed fluctuations magnetic field compressibility, $C_{b_i}$ as defined in  \cite{brubav1991} as: $\sigma_{B}$/$\sigma_{b_i}$ with $\sigma_B$ and $\sigma_{b_i}$ the variance of magnetic field magnitude and fluctuations (with i = x, y, z), respectively.
A lower compressibility is observed during the Alfv\'enic periods. This is in agreement with the results found by \cite{damicis2015}. Actually, those authors showed that the highest Alfv\'enicity values, quantified by high values of the normalized cross-helicity, are coupled to the lowest values of field compressibility. This statistical study shows that solar wind fluctuations during the maximum of cycle 23 were less affected by compressive phenomena and, as a consequence, the maximum of this cycle was more Alfv\'enic than its corresponding minimum.

The Coulomb collisional age ($A_C$) is a measure of the efficacy of Coulomb
collisions, estimating the number of binary collisions in each plasma parcel during transit from the Sun to the spacecraft. The collisional age for protons
was estimated according to \cite{kasper2008} as:
$A_C$ = $\nu_{pp} R/V_{sw}$ with $\nu_{pp} \propto n_p T^{-3/2}$
where $\nu_{pp}$ is the proton-proton collision frequency while $R/V_{sw}$ is the transit time from the Sun to the s/c.
A larger $A_C$ means that the plasma has
undergone more Coulomb collisions, while low values of $A_C$
are associated with plasmas that experienced fewer Coulomb
collisions and thus can be expected to better preserve any
signatures of processes experienced in the inner corona.
Figure \ref{figure3} shows that the highest $A_C$ values correspond to the typical slow wind, oscillating around $10^{-1}$. The two Alfv\'enic winds on the contrary are less collisional even if they are not characterized by the same average $A_C$ values. Actually, the Alfv\'enic slow wind is on average slightly higher than $10^{-2}$ while the fast wind values are well below that value.  

Thermal speed, $V_{th} = (2 k_B T_p /m_p)^{1/2}$, and Alfv\'en speed, $V_A = B/(\mu_0 n_p m_p)^{1/2} $, are displayed Figure \ref{figure3} as well. The bottom panel displays the plasma beta, $\beta$ obtained as $V_{th}^2/V_A^2$. Most of the time interval under study is characterized by a higher thermal speed than the Alfv\'en speed especially within the Alfv\'enic slow solar wind. The typical slow wind shows a similar behaviour even if a large decrease of the Alfv\'en speed is observed in correspondence of the decrease in the B magnitude observed in Figure \ref{figure1}, most likely linked to the presence of a magnetic hole, term introduced by \cite{turner} and later better defined as magnetic decreases \citep[][]{tsuru2002a,tsuru2002b,tsuru2011a}, occurred during most of the day 34 \citep[see for example also][]{winterhalter,franz}. Actually, in the Alfv\'enic slow wind, $V_A$ is always larger than $V_{th}$ determining a $\beta$ always less than one. During the typical slow wind, in some small intervals these two quantities are almost equal determining a $\beta$ close to 1. In the fast wind case this behaviour is even clearer, with $V_{th}$ typically equal to $V_A$ and a consequent $\beta$ around 1 during the entire fast stream.  

\section{Microphysics}

To address the properties of particle microphysics, it is useful to use a parameter space which combines the (parallel) plasma beta of a given species to its temperature anisotropy \cite[e.g.][]{gary2002}. In this parameter space, the thermodynamical state of the plasma can be tested against kinetic instabilities, such as mirror and ion-cyclotron (generated when $T_\perp > T_\parallel$) and fire hoses ($T_\perp < T_\parallel$) \citep[e.g.][]{hellinger, bale2009}.
Typical fast and slow wind plasma display different distributions in this parameter space.
Moreover, \cite{marsch2004} have revealed the presence of a linear relation between the parallel proton beta $\beta_{\parallel}$ and the core temperature anisotropy $T_\perp/T_\parallel$ in fast streams. Further studies have confirmed such a relation for total temperatures and \cite{matteini2007} have interpreted this behavior as an evolutionary path of the plasma with radial distance, starting with $T_\perp > T_\parallel$ and low proton beta close to the Sun, and then progressively approaching the fire hose unstable region at larger distances (higher $\beta$). The same path/relation with radial distance is not observed in typical slow wind intervals, where the distribution of the observational counts appears to be bounded by the thresholds of kinetic instabilities at all distances.
It is then interesting to test the properties of the Alfv\'enic slow plasma in the ($\beta_{\parallel},T_\perp/T_\parallel$) space, to check whether its similarities with fast streams extends also to the microphysics.

\begin{figure}
	\includegraphics[width=250pt]{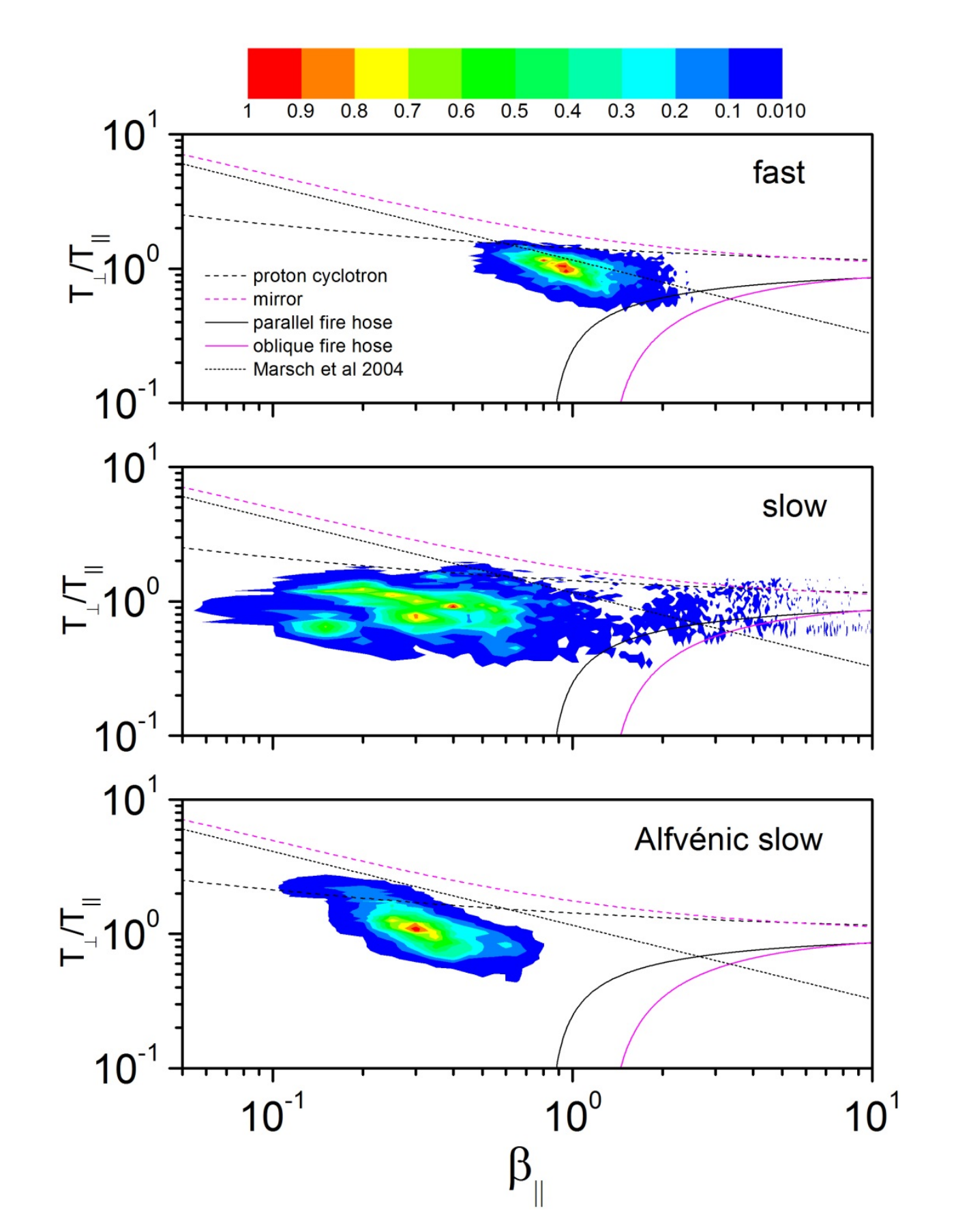} 
	\caption{Contour plots showing the relationship between $\beta_{\parallel}$ and anisotropy $(T_{\perp}/T_{\parallel})$ for the three solar wind regimes. In each plot the main plasma instability threshold are displayed: proton cyclotron (black dashes), mirror (red dashes), parallel fire hose (black line), oblique fire hose (red line), relation by \citet{marsch2004} (black dots).}
	\label{figure4}
\end{figure}

Figure~\ref{figure4} shows the distribution of the proton anisotropy as a function of $\beta_{p\parallel}$ for the case study already presented in the previous sections representing different solar wind regimes.
The top panel shows fast wind distribution which is well-centered around $\beta_\parallel$ and $T_\perp/T_\parallel$. Moreover it follows quite well the relationship $T_\perp/T_\parallel \sim 1.16 \beta_{p \parallel}^{-0.55}$ found by \cite{marsch2004}. The middle panel displays a typical slow wind distribution. It is characterized by spread values in $\beta_\parallel$ while anisotropy values are centered around 1. The bottom panel represents the case for the Alfv\'enic slow wind. It must be noted in this case that although an anti-correlation still exists between anisotropy and $\beta$, this distribution does not follow very closely the relationship by \cite{marsch2004} rather it is found to be almost parallel to it since it has lower $\beta$ values and larger anisotropy values (at least for the maximum values).  

It is quite clear from these plots that also in terms of the proton microphysics (level of anisotropy, correlation with  $\beta_{p\parallel}$, and relation with kinetic instabilities) the Alfv\'enic slow wind is very close to fast wind observations, and shows a different thermodynamical state with respect to the typical slow wind, whose anisotropy is weakly correlated with beta.

\section{$\theta_{BR}$-$V_{SW}$ relationship}

As mentioned thoughout this paper, fast streams are characterized by a high level of Alfv\'enicity, namely the high degree of correlation between velocity and magnetic field fluctuations. Related to this state is also the constancy of magnetic and plasma pressures, keeping the fast wind in a regime of low compressibility.

\begin{figure}
	\centerline{\includegraphics[width=250pt]{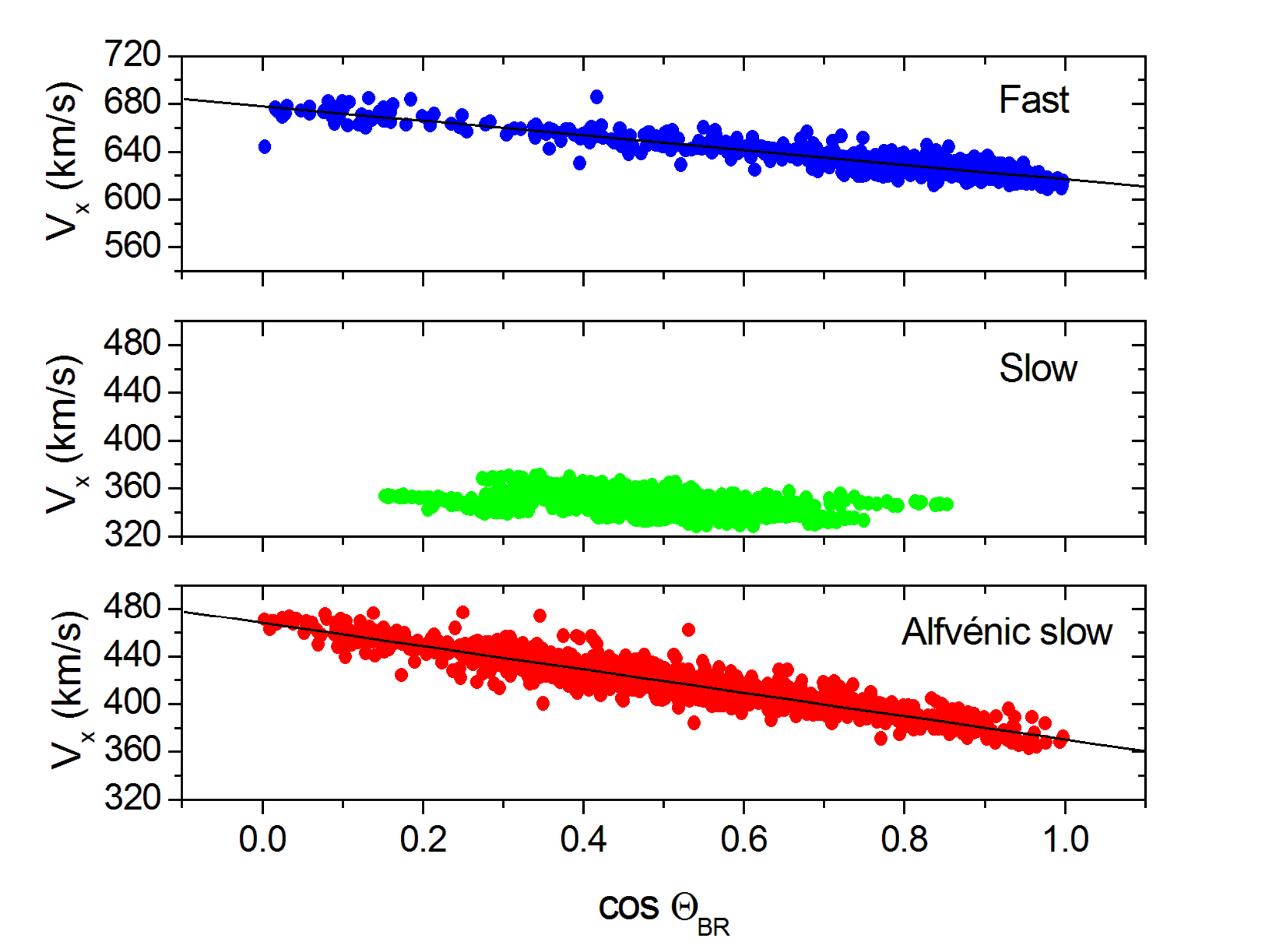}} 
	\caption{Relationship between $\theta_{BR}$ and $V_{sw}$ for fast and Alfv\'enic slow solar wind.}
	\label{figure5}
\end{figure}

On the other hand, the typical slow wind, is much more irregular and display large variations of magnetic, gas, and total pressures, together with a much weaker Alfv\'enic correlation.
It has been recently shown \citep{matteini2014, matteini2015} that a consequence of the high-Alfv\'enicity state of the fast solar wind is the modulation of the flow bulk speed by the direction of the local magnetic field. In other words, there is a well-defined  correlation between the proton speed and the cosinus of angle $\Theta_{BV}$ between the instantaneous magnetic field and the solar wind velocity \citep{matteini2014}.
The top panel of Figure \ref{figure5} shows such a correlation for the fast wind interval. As expected, data align along a straight line, whose slope approximatively corresponds to the phase velocity of the Alfv\'enic fluctuations which can be inferred by inspection of the correlation between magnetic and  velocity fluctuations. It is known that the empirical phase velocity can deviate slightly from the nominal Alfv\'en speed due to the presence of residual energy in the plasma (i.e. excess of magnetic to kinetic fluctuating energy).
\cite{matteini2014} derived the empirical relation:
\begin{equation}\label{eq_thetabv}
V_p\sim V_0+v_{ph}[1-\rm{cos}(\theta_{BR})]
\end{equation}
where $V_0$ the minimum speed within the interval and $v_{ph}$ takes into account the observed residual energy $R_A$, or Alfv\'en ratio, so that $v_{ph}=V_A\sqrt{R_A}$. For the sample of fast wind  $r_A$ is 0.61 and $V_A$ = 71 km s$^{-1}$ giving a value of 56 km s$^{-1}$ which should be compared with the slope of the fit which corresponds 61 km s$^{-1}$.

For the typical slow wind a linear correlation is not expected, since not Alfv\'enic, and indeed the central panel shows that the speed has not a clear dependence on the angle $\Theta_{BV}$.
On the contrary the Alfv\'enic slow wind interval, displayed in the bottom panel, does show the same correlation as the fast wind, and again, this is very well reproduced by relation \ref{eq_thetabv}. In this case $r_A$ is 0.89 and $V_A$ = 111 km s$^{-1}$ giving a value of 104 km s$^{-1}$ which should be compared with the slope of the fit which corresponds 98 km s$^{-1}$.

This additional test of the fluid behavior of the Alfv\'enic slow wind highlights further how its properties are very similar to the fast wind from coronal holes, and that the broad phenomenology characterising fast streams is fully recovered also on slower streams that are equally Alfv\'enic.

\section{Spectral properties}

The characteristics described in the previous sections clearly show similarities between the Alfv\'enic slow wind and the fast wind. This has implication on the spectral features as well. 

Solar wind fluctuations show a typical Kolmogorov-like power law in the inertial range. In particular, in fast wind, at low frequencies, a $k^{-1}$ scaling is clearly present. The origin of this $k^{-1}$ scaling is still highly debated \citep[e.g][submitted]{mg1986,dmitruk2007,verdini2012,tsuru2018,matteini}.
A frequency break separates the $k^{-5/3}$ from the $k^{-1}$ scaling. This break corresponds to the correlation length.

At kinetic scales energy is eventually dissipated. As a matter of fact, around the proton scales, another spectral break is found beyond which the spectrum generally steepens, with different slopes for fast and slow wind \citep{bruno2014,smith2006,sarahoui}. 

\begin{figure}
	\centerline{\includegraphics[width=250pt]{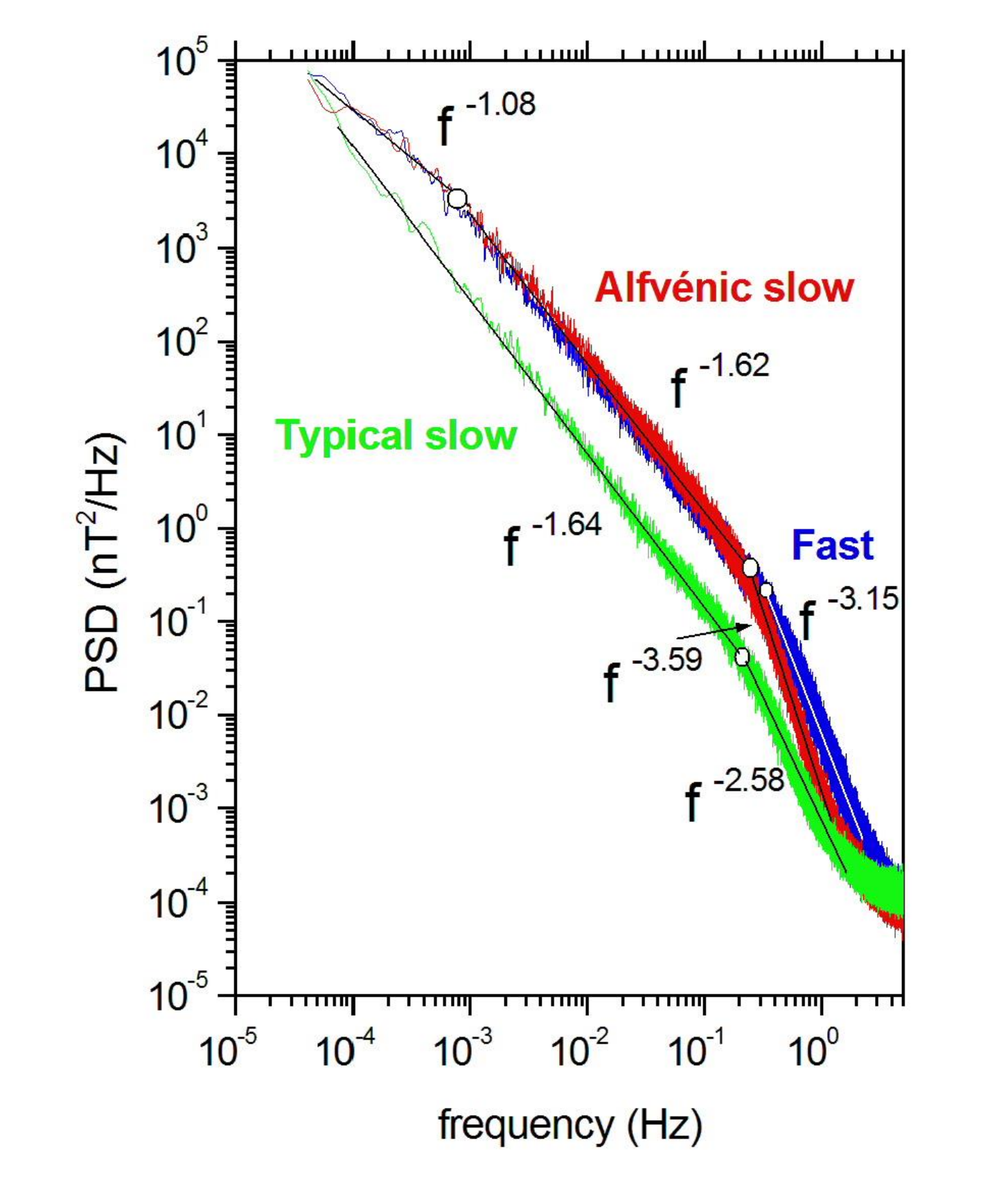}} 
	\caption{Trace of the power spectral density of magnetic field fluctuations up to the sub-ion range for typical slow (green), Alfv\'enic slow (red) and fast (blue) wind, respectively. Superimposed to the power spectra the fit in each frequency range are shown along with the position of the spectral breaks separating the different frequency domains.}
	\label{figure6}
\end{figure}

Figure~\ref{figure6} shows an example of power spectral density of magnetic field fluctuations for the three solar wind regimes under study from small to large frequencies, using data from the MFI experiment on board WIND \citep{lepping1995} at 92 ms resolution. 

This part of the analysis is based on the study of eight intervals for each solar wind regimes introduced throughout the paper chosen with the following criteria and as indicated in table \ref{tab:summary}. Fast wind and typical slow wind where chosen during minimum of solar activity, during years 1995 and 2005, since in this phase of the solar cycle the fast streams usually are recurrent streams, within which the main portion of the fast stream and the rarefaction region can be easily identified. 
However, the characteristics highlighted in the first part of this paper are found also in fast and typical slow wind at minimum of solar activity.
The eight intervals of Alfv\'enic slow wind instead were selected during maximum of solar cycle 23. Table \ref{tab:summary} contains the main average parameters that characterize these time intervals and in particular: solar wind bulk speed, Alfv\'en speed, thermal speed, number density, magnetic field magnitude.     

\begin{table*}
	\caption{Time intervals referring to fast (F), typical slow (TS) and Alfv\'enic slow (AS) solar wind observed with WIND showing the following parameters: solar wind bulk speed (V$_{sw}$), Alfv\'en speed (V$_A$), thermal speed (V$_{th}$), proton number density (n$_p$), magnetic field magnitude (B), cyclotron frequency ($\Omega_p$), frequency break ($f_b$), angle between magnetic field and radial direction ($\theta_{BR}$), plasma beta ($\beta$). The other characteristics lenghts can be derived accordingly. \label{tab:summary}}
	\begin{tabular}{cccccccccccc}
	\hline
	\hline
		\# & Year & Time interval & V$_{sw}$ & V$_A$ & V$_{th}$ &
		n$_p$ & B & $\Omega_p$ & $f_b$ & $\theta_{BR}$ & $\beta$ \\
		&	& 	DoY:hh & km s$^{-1}$ & km s$^{-1}$ & km s$^{-1}$ &
		cm$^{-3}$ & nT & Hz & Hz & $\deg$ & \\
	\hline
	1F & 1995 & 003:20 - 004:16 & 686.6 & 68.7 & 78.6 & 3.51 & 5.90 & 0.090 & 0.39 & 51.0 & 1.31 \\
	2F & 1995 & 030:12 - 031:12 & 708.6 & 74.9 & 81.3 & 2.70 & 5.64 & 0.086 & 0.36 & 13.8 & 1.18\\
	3F & 1995 & 098:00 - 099:00 & 666.1 & 64.3 & 74.9 & 3.24 & 5.31 & 0.081 & 0.35 & 41.5 & 1.35 \\
	4F & 1995 & 126:00 - 127:00 & 726.2 & 63.5 & 80.9 & 2.83 & 4.90 & 0.074 & 0.41 & 12.8 & 1.62\\
	5F & 2005 & 067:00 - 068:00 & 744.7 & 75.5 & 83.2 & 2.62 & 5.60 & 0.085 & 0.35 & 33.2 & 1.22 \\
	6F & 2005 & 282:00 - 283:00 & 655.2 & 77.4 & 75.4 & 1.85 & 4.83 & 0.073 & 0.28 & 43.0 & 0.95\\
	7F & 2005 & 308:12 - 309:12 & 716.6 & 90.3 & 81.4 & 1.44 & 4.97 & 0.076 & 0.31 & 46.3 & 0.81\\
	8F & 2005 & 335:00 - 336:00 & 736.5 & 96.5 & 85.3 & 2.00 & 6.26 & 0.095 & 0.35 & 41.7 & 0.78 \\
	\hline
	1TS & 1995 & 009:00 - 010:00 & 469.3 & 49.9 & 47.3 & 3.96 & 4.42 & 0.067 & 0.26 & 21.0 & 1.10 \\
	2TS & 1995 & 036:01 - 036:08 & 477.1 & 50.2 & 47.8 & 2.81 & 3.83 & 0.058 & 0.26 & 39.4 & 0.98 \\
	3TS & 1995 & 104:04 - 105:04 & 444.8 & 36.4 & 43.7 & 4.08 & 3.34 & 0.051 & 0.20 & 37.9 & 1.78 \\
	4TS & 1995 & 130:12 - 131:12 & 428.7 & 41.1 & 43.8 & 4.84 & 4.08 & 0.062 & 0.28 & 5.1 & 1.90 \\
	5TS & 2005 & 071:00 - 072:00 & 410.6 & 38.5 & 38.9 & 2.42 & 2.71 & 0.041 & 0.15 & 23.2 & 1.25 \\
	6TS & 2005 & 287:04 - 287:18 & 366.8 & 42.2 & 34.6 & 3.71 & 3.61 & 0.055 & 0.20 & 30.9 & 0.78 \\
	7TS & 2005 & 313:16 - 314:00 & 440.6 & 54.2 & 44.0 & 1.73 & 3.21 & 0.049 & 0.21 & 44.8 & 0.97 \\
	8TS & 2005 & 341:12 - 342:00 & 374.1 & 48.8 & 34.1 & 1.60 & 2.81 & 0.043 & 0.23 & 17.7 & 0.52 \\
	\hline
	1AS & 2000 & 351:12 - 352:12 & 368.4 & 48.4 & 43.8 & 7.24 & 5.96 & 0.091 & 0.24 & 54.1 & 0.82 \\
	2AS & 2001 & 040:00 - 041:00 & 426.8 & 54.8 & 47.9 & 4.49 & 5.32 & 0.081 & 0.25 & 57.9 & 0.76 \\
	3AS & 2001 & 067:00 - 068:00 & 469.3 & 57.9 & 48.9 & 3.71 & 5.12 & 0.078 & 0.28 & 47.6 & 0.71 \\
	4AS & 2001 & 327:00 - 328:00 & 462.8 & 73.1 & 62.9 & 3.87 & 6.59 & 0.100 & 0.26 & 27.3 & 0.74 \\
	5AS & 2001 & 340:00 - 341:00 & 441.9 & 58.4 & 52.0 & 5.82 & 6.46 & 0.098 & 0.26 & 35.8 & 0.79 \\
	6AS & 2001 & 352:00 - 353:00 & 472.5 & 67.7 & 51.9 & 2.83 & 5.22 & 0.079 & 0.23 & 44.4 & 0.59 \\
	7AS & 2002 & 026:00 - 027:00 & 416.5 & 100. & 60.1 & 3.64 & 8.77 & 0.133 & 0.26 & 55.8 & 0.36 \\ 
	8AS & 2002 & 056:00 - 057:00 & 337.0 & 54.5 & 36.3 & 6.84 & 6.53 & 0.099 & 0.20 & 46.4 & 0.44 \\
	\hline
	\end{tabular}
\end{table*}

When moving from fast to typical slow wind one observes a different power level, higher in the fast wind rather than in the typical slow wind indicating the presence of larger magnetic field fluctuations. Interesting enough is that the same power characterizes also the Alfv\'enic slow wind. Moreover, a spectral break separating the inertial range from the injection range, identifying the correlation length, is located at the same frequency at about 30 min, which is a timescale typical of Alfv\'enic fluctuations. This corresponds to a spatial scale of about $0.8 \cdot 10^6$ km and of $1.2 \cdot 10^6$ km, for Alfv\'enic slow and fast wind respectively, being in accordance with previous studies \citep[e.g.][etc.]{mg1982,brunodobro,matthaeus2005}.

The typical slow wind, on the contrary, is generally characterized by a Kolmogorov scaling extending to much lower frequencies as recently reported \citep{rb2017}. 
However, a detailed discussion on the low frequency part of the spectrum is out of the scope of our paper so that the interested reader can refer to the above bibliographic references for further reading.

Besides the differences at low frequencies, there are also clear differences at proton kinetic scales. \cite{bruno2014}, investigating the behavior
of the spectral slope at proton scales, up to frequencies of a few Hz, beyond the high-frequency break separating fluid from kinetic scales, recorded a remarkable variability of
the spectral index \citep{bruno2014,smith2006,sarahoui}
within the following frequency decade or so. The steepest spectra correspond to the main portion of fast streams while the flattest ones were found within the subsequent slow wind regions \citep{bruno2014}. This is confirmed by our findings shown in Figure \ref{figure6}. In addition, our analysis clearly shows that the kinetic break of the typical slow wind is located at lower frequencies than the two Alfv\'enic winds. It must be noted, however that, although the behaviour of the Alfv\'enic slow wind is similar to the fast wind, its kinetic break is a bit shifted towards lower frequencies.  

\cite{bruno2014} found also a strong dependence of the observed spectral
slope at ion scales and the power characterizing the
fluctuations within the inertial range: the higher the power, the
steeper the slope. This parameter shows the steepest spectra within the main portion of the fast streams, where the speed is higher, and the lowest
values within the subsequent slow wind, following a gradual
transition between these two states. The dependence of the spectral slopes in the dissipation range
on the power level in the corresponding inertial range is rather robust since the same kind of relationship
applies equally well to data points belonging to different time
intervals.

\begin{figure}
	\includegraphics[width=200pt]{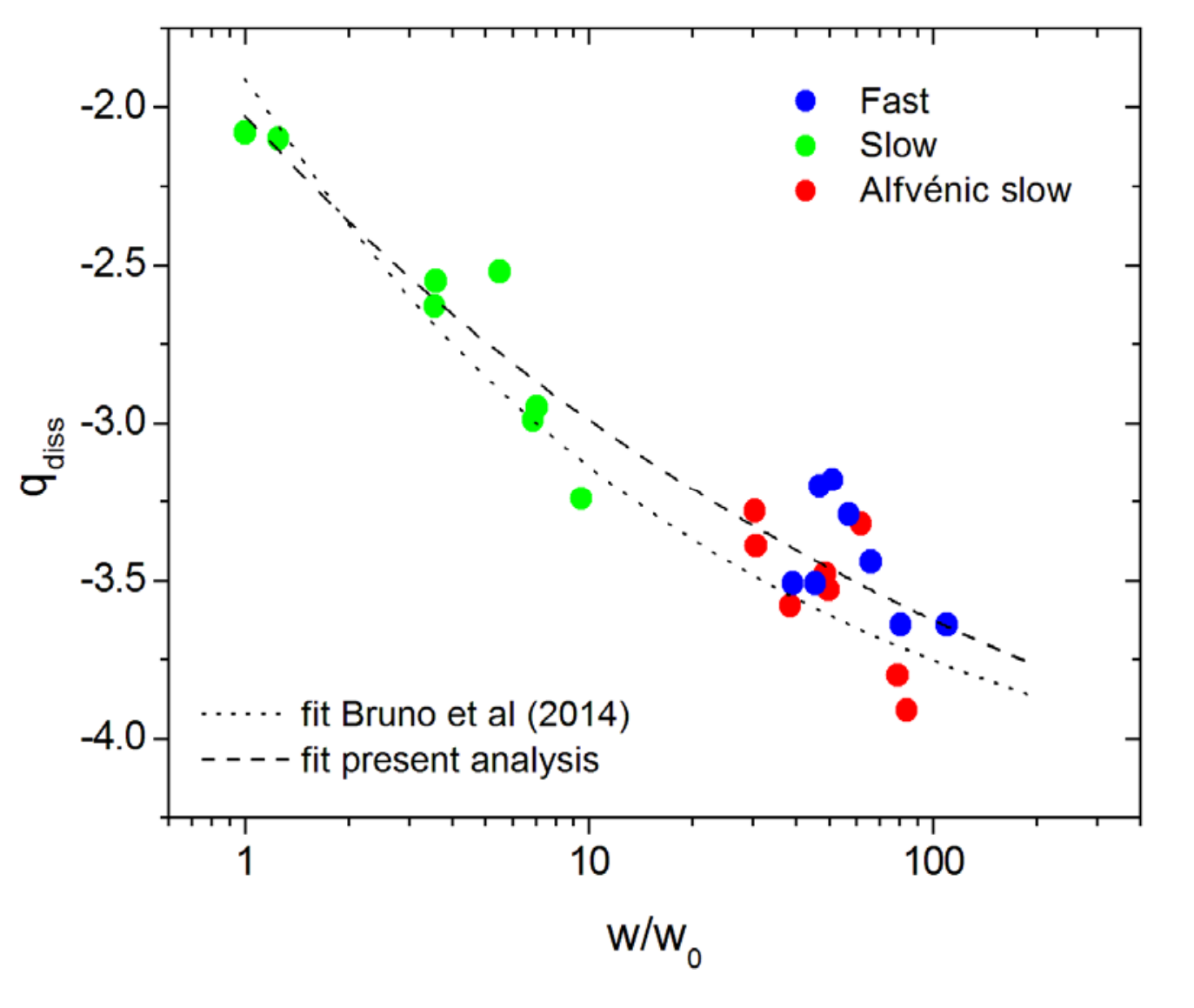} 
	\caption{Dependence of the spectral index in the dissipation range ($q_{diss}$) in the sub-ion range on the power associated with the fluctuations within the inertial range ($w/w_0$), where $w_0$ is normalized to the lowest power spectrum. The best fit is shown in comparison to the one by \citet{bruno2014}. The two fits overlap with error bars with 95\% confidence level.}
	\label{figure6bis}
\end{figure}

Figure~\ref{figure6bis} shows the dependence of the spectral index in the dissipation range ($q_{diss}$)in the sub-ion range on the power associated with the fluctuations within the inertial range ($w/w_0$), where $w_0$ is normalized to the lowest power spectrum for the present study.
The power associated to the inertial range was computed as the integral of the
PSD in a frequency range chosen within the inertial range in the same range as in 
\cite{bruno2014}, namely from $7 \times 10^{-3}$ to $10^{-1}$ Hz. We
normalized the values of the integrated PSD to the lowest
power ($w_0$) in the inertial range which corresponds to a typical low-speed wind interval 
in order to have a dimensionless parameter on the X axis.

The spectral slopes in the dissipation range were obtained
through a fitting procedure, taking care to exclude regions
too close to the break point or at higher frequencies where the
spectrum flattens \citep[e.g.][]{brunotrenchi2014}.
The best fit q = (-4.86 $\pm$ 1.28)+(2.83 $\pm$ 1.21) ($w/w_0)^{(−0.18 \pm 0.13)}$
was obtained using a power-law fit, shown by a dashed
black line where q is the spectral index and $w/w_0$ indicates the normalization
process performed within the inertial range. It is interesting
to note that this fit is quite in agreement with the one found in 
\cite{bruno2014}. The latter is shown for reference and it is found to be in good accordance with the data of the present analysis. The two fits overlap with error bars with 95\% confidence level.

\section{Location of the kinetic break}

In the previous section we have discussed the general behaviour of the power spectra in terms of power level and slope in the different frequency regimes with a quick look at the break separating the $k^{-5/3}$ from the $k^{-1}$ scaling.

Another important issue is found in the location of the spectral break at kinetic scales addressed by several authors \citep[see reviews by][]{alexandrova2013,brunocarbone2013} but still highly debated. 

At these scales, protons are continuously heated during the wind expansion \citep{marsch2012}. One possible source of proton heating is represented by some form of dissipation, at the proton kinetic scale, of the energy
transferred along the inertial range. This would change the scaling exponent. There are different relevant lengths which can be associated with this phenomenon, depending on the particular dissipation mechanism we consider as widely discussed in \cite{brunotrenchi2014,sarahoui,leamon1998,markovskii2008,marsch2006,gary1993,chen} so different characteristic lengths indicate a different dissipation mechanism invoked to explain the observed local heating of the solar wind plasma.

The characteristic scales which might correspond to the observed spectral break are the proton inertial length $\lambda_i = c/\omega_p$ and the proton Larmor radius $\lambda_L = v_{th}/\Omega_p$, expressed in cgs units. $\omega_p = (4 \pi n_p q^2/m_p)^{1/2}$ and $\Omega_p = q B/(m_p c)$ are the plasma and cyclotron frequencies, respectively, where q is the proton electric charge, n$_p$ the proton number density, B the local magnetic field intensity, m$_p$ the proton rest mass, and c the speed of light. Since c/$\omega_p = v_A/\Omega_p$, the proton inertial length can also be expressed as $\lambda_i = v_A/ \Omega_p$, where v$_A$ is the Alfv\'en speed. 

Although turbulence phenomenology would limit the role played by parallel
wavevectors k$_{\parallel}$, \cite{brunotrenchi2014} firstly found that
the cyclotron resonant wavenumber showed the best agreement with the location of the break compared to results obtained for the ion inertial length and the Larmor radius, result confirmed also by findings in \cite{telloni2015}.

\cite{chen} investigated extremely low $\beta$ (about $10^{-2}$) and extremely high $\beta$ (about 20) intervals finding that the ion inertial length scale plays a role in the location of the high-frequency break for low beta plasma while the ion gyroradius for high plasma beta. On the other hand, although these authors observed also that the cyclotron resonant wave number does fit the observations at both high and low $\beta$ within errors, they did not consider a possible role of k$_{\parallel}$ because it was not consistent with the observed anisotropy of turbulence \citep{horbury2008}.

In clear support to the resonance condition hypothesis, a recent study by \cite{woodham2018} found that the high-frequency spectral steepening is best associated with the cyclotron resonance scale, both in fast and slow wind streams, as well as periods where $\beta \sim 1$ where the agreement is strongest.

The two groups with different and extreme plasma $\beta$ described in \cite{chen} have been harmonized by a recent study by \cite{wang2018} who investigated ion-scale spectral break frequencies over a full-range plasma beta. They found results in support of the cyclotron resonance condition playing a vital role in the dissipation process at the spectral break at all beta. By analyzing the normalized frequency breaks, they found that the average value of $f_b/f_{R}$ (frequency break normalized to the frequency corresponding to the resonance condition) in each $\beta$ bin seems to be nearly a constant and not dependent on $\beta$. It was also found that the ratio between $f_b$ and $f_{R}$ is statistically close to 1. For intermediate $\beta$, $f_b/f_{R}$ is much closer to 1 than both $f_b/f_{i}$ (frequency break normalized to the frequency corresponding to ion inertial length) and $f_b/f_{L}$ (frequency break normalized to frequency corresponding to ion gyroradius). But when $\beta \ll 1$ ($\beta \gg 1$), $f_b/f_{i}$ ($f_b/f_{L}$) is also nearly unity.

In the present study the kinetic break was obtained in the frequency domain as the intersection between the fits computed within the inertial range and the kinetic range. For each frequency range a fitting procedure was applied, taking care to exclude regions too close to the break point. The frequency break $f_b$ was then transformed in wavenumber $k_b$ taking into account that $k_b = 2 \pi f_b/ V_{sw}$ with $V_{sw}$ the solar wind bulk speed. Also the quantities $\lambda_L$ and $\lambda_i$ were computed in terms
of wavenumber $k$ as $k_L$ = $\Omega_p/V_{th}$ and $k_i = \Omega_p/V_A$ while the resonant condition can be expressed as $k_R = \Omega_p/(V_A + V_{th})$. For sake of completeness, we considered also $k_C$ = $2 \pi \Omega_p/V_{sw}$ associated to the cyclotron frequency $\Omega_p$ as a reference. All these characteristic lengths can be derived from the parameters shown in Table~\ref{tab:summary}. 

\begin{figure}
	\centerline{\includegraphics[width=250pt]{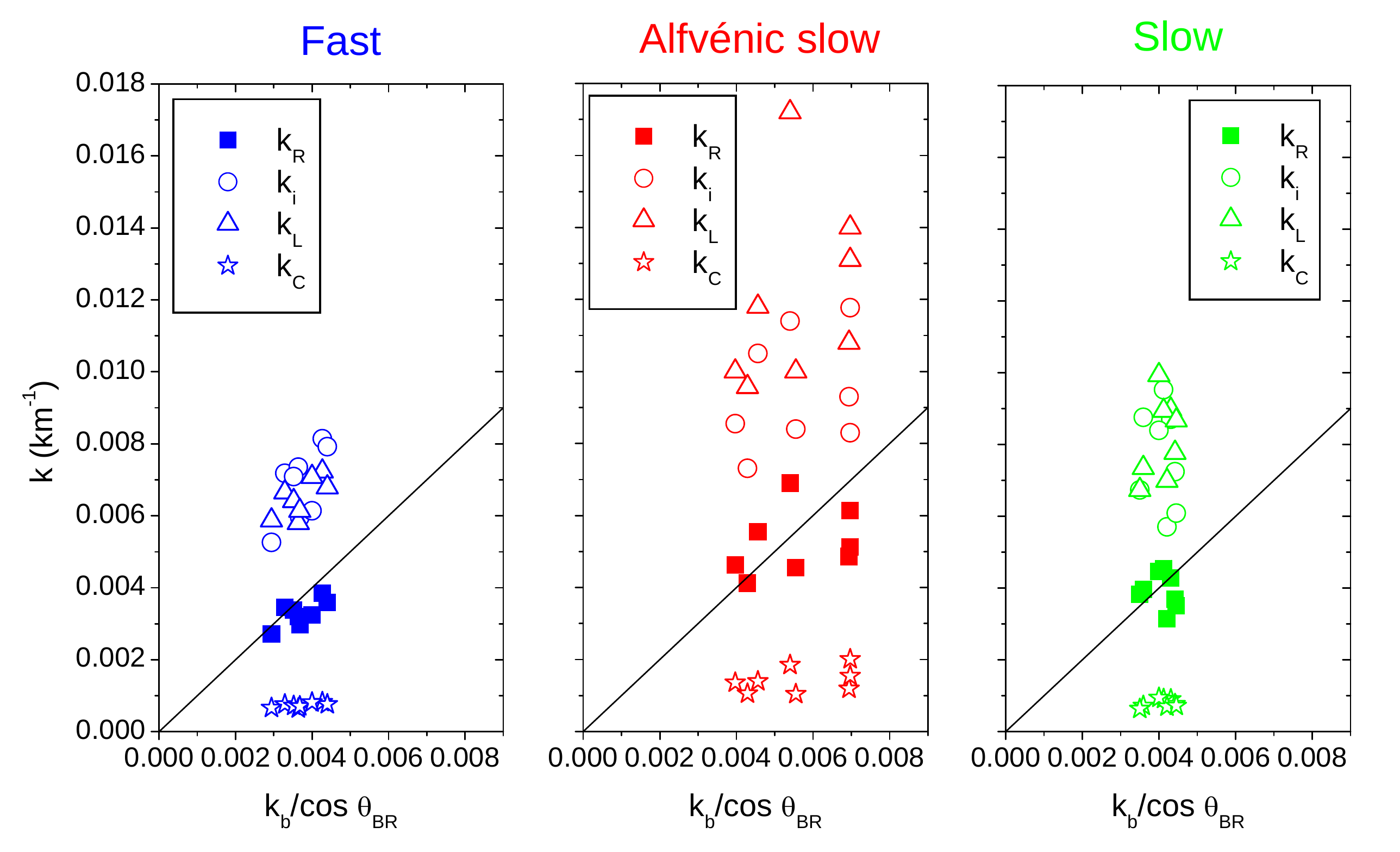}} 
	\caption{Location of the spectral break, in terms of wavenumbers, normalized to cos $\theta_{BR}$ vs some plasma characteristic lengths: resonance condition $k_R$ (full squares), inertial length $k_i$ (empty circles), Larmor radius $k_L$ (empty triangles), $k_C$ associated to the cyclotron frequency (empty stars). The color code corresponds to the different solar wind regimes: green (typical slow), blue (fast), red (Alfv\'enic slow).}
	\label{figure7}
\end{figure}

We then computed for each solar wind regime under study the characteristic lengths along with the position of the break normalized to $\cos \theta_{BR}$ (indicated for each interval in Table~\ref{tab:summary}) to take into account that $k$ is along the direction of the local mean field
while we are sampling along the radial direction at
an angle $\Theta_{BR}$. Our results are displayed in Figure~\ref{figure7} in a similar way to what performed by \cite{brunotrenchi2014}.

Our study confirms that the best agreement for each solar wind regime is found for the wavenumber $k_R$ corresponding to the resonance condition for parallel propagating Alfv\'en waves. $k_i$ and $k_L$ are always much larger that $k_R$ while $k_C$ is always lower. Regarding the comparison of the position of the break, $k_b$ is found at larger values for the Alfv\'enic slow solar wind (scales around 140-220 km), at intermidiate values for the typical slow solar wind (scales around 250 km), at small-intermidiate values for the fast wind (scales around 220-330 km).  

Table~\ref{tab:summary} contains also the kinetic frequency break and the plasma $\beta$ for the selected time intervals studied. Our study shows that $k_R$ fits better all beta although for the selected intervals this parameter ranges between 0.36 and 1.78.   

\section{Summary and conclusions}

This study shows that the standard classification of the solar wind in two categories according to their speeds should be revised as suggested also by e.g. \cite{vonsteiger2008,zhao2009,stakhiv2015,xu2015,camporeale2017,ko2018,stansby}. For instance, the Alfv\'enic content of solar wind flcutuations is different when moving from fast to slow streams with fast wind more Alfv\'enic than slow wind. 
However, a recent result has shown that also the slow solar wind can sometimes show a high degree of Alfv\'enicity \citep{damicis2015,damicis2016} in contrast with previous findings. Although the existence of this kind of slow wind was first pointed out in the 80s but close to the Sun, a detailed characterization was not developed at that time. The added value of the present paper is threefold: it gives a comprehensive characterization of this kind of slow solar wind; it uses measurements at 1 AU where we expect a depletion of Alfv\'enic correlations; this kind of solar wind was found on a statistical basis at maximum of solar cycle 23. These are strong arguments which prompt our study. This Alfv\'enic slow solar wind is thus the focus on the present study.

To compare the different solar wind regimes, in the first part of this paper, we show a limited but representative time interval of about 20 days. The investigation we carried out takes into account different aspects of the physics of the interplanetary medium but every aspect demonstrates that the Alfv\'enic slow wind resembles the fast wind rather than the typical slow wind. Here is a summary of the main findings. 

The Alfv\'enic slow wind has a bulk speed comparable to the typical slow wind. However, it is characterized by larger fluctuations in both velocity and magnetic field components (not shown here) typical of the presence of Alfv\'enic fluctuations. Actually, Alfv\'enic periods are usually well detectable also by looking at the velocity profile where fluctuations are enhanced respect to periods of low Alfv\'enic correlations. Overall the bulk parameters show a clear similarity with the fast wind features. For instance constant magnetic field magnitude and number density indicate low compressibility typical of this kind of fluctuations. However, the bulk speed, besides being lower, is almost constant and does not show a profile typical of the fast wind stream (i.e. presence of main portion of the stream and a rarefaction region).  

Regarding temperature, the Alfv\'enic slow wind is hotter than the typical slow wind but colder than the fast wind, following the well-known temperature-speed relation.

Related to the previous point, a lower compressibility, as defined in \cite{brubav1991}, is observed during Alfv\'enic periods thus supporting 
the idea that compressible phenomena may act in destroying the Alfv\'enicity of solar wind fluctuations. 
This is also in agreement with the results by \cite{damicis2015} who found that lowest values of field compressibility characterize the maximum of solar cycle 23 and were then coupled with a higher Alfv\'enic content than its corresponding minimum. 

The Alfv\'enic intervals are characterized by lower O$^{7+}$/O$^{6+}$ and C$^{6+}$/C$^{5+}$ respect to the typical slow wind. Actually, the Alfv\'enic populations fill the same region in a scatter plot O$^{7+}$/O$^{6+}$ vs C$^{6+}$/C$^{5+}$ and are distint from the typical slow wind. This finding supports the idea of a common solar origin of the two kinds of Alfv\'enic winds, hypothesis supported by recent findings by \cite{damicis2015}.  

The two Alfv\'enic winds show lower values of $A_C$, associated with plasmas that experienced fewer Coulomb collisions and thus expected to better preserve any signatures of processes experienced in the inner corona.

A very interesting difference is the behaviour of $\beta$ in the different solar wind regimes under study. We performed a study, not shown here graphically, in which we studied the dependence of $\beta$ from thermal and Alfv\'en speed and found that the Alfv\'enic slow wind is characterized by a higher Alfv\'en speed respect to the thermal speed and it is then always characterized by $\beta$ less than 1. In the intervals of fast and typical slow wind wind, on the contrary, thermal and Alfv\'en speed are comparable so that $\beta$ is on average around 1. In general, however, $V_A$ and $V_{th}$ are lower in the Alfv\'enic slow rather than in the fast wind.   

We studied also the properties of the solar wind plasma in the ($\beta_{\parallel},T_\perp/T_\parallel$) space, comparing the different solar wind regimes. We found that the similarities between fast streams extend also to the microphysics. Actually, the fast wind distribution follows quite well the relationship by \cite{marsch2004}. The Alfv\'enic slow wind shows similar behavior but it is found at lower $\beta$ values and larger anisotropy values (at least for the maximum values).

This study also confirms previous papers by \cite{matteini2014, matteini2015} finding, for the Alfv\'enic winds, a well-defined correlation between the proton speed and the cosinus of angle $\Theta_{BV}$ between the instantaneous magnetic field and the solar wind velocity, further supporting the idea that the high-Alfv\'enicity state of the fast solar wind determines the modulation of the flow bulk speed by the direction of the local magnetic field. 

In the second part of this paper dedicated to the study of the spectral features, we performed a more quantitative analysis considering different intervals for each solar wind regime.  

The comparison between the power spectra shows that the two Alfv\'enic winds are characterized by higher power related to a larger amplitude of the magnetic field fluctuations which are typical of the presence of Alfv\'enic fluctuations. The lower frequencies are characterized also by a break separating the inertial range from the larger scales with a typical scale of tenths of minutes. The presence of this break might be related to the turbulent age of fluctuations at a given scale: the faster is the wind or the stronger is the Alfv\'enic correlation than the younger is the turbulence. Since the coronal spectrum is supposed to be rather flat (at least in the fast solar wind), smaller spectral indices correspond to less evolved spectra. According to this interpretation, one would expect spectral slope to change with distance as the turbulence ages, while observations report fairly stable spectral slopes. Although we performed a preliminary analysis on this relationship, we will postpone a detailed investigation to a future paper. 

This study further confirms that the slope of the spectrum in the dissipation range in the sub-ion range depends on the power of the fluctuations in the inertial range: a steeper spectrum corresponds to a higher power in agreement with \cite{bruno2014}. With this respect, the two Alfv\'enic winds are characterized by similar power and slopes and have larger values respect to the typical slow wind. This has been done considering several time interval to better support our findings. 

A highly debated topic is the location of the kinetic break which can help in identifying the dissipation mechanism responsible for the local heating of the solar wind plasma.  \cite{brunotrenchi2014} first found the best agreement of the resonant wavenumber with the location of the kinetic break rather than with the ion inertial length and the Larmor radius. This result is then strongly supported also by more recent papers \citep{woodham2018, wang2018} and by our study. Actually, Figure \ref{figure7} shows that $k$ associated with the resonance condition fits better the kinetic break rather than other characteristic lengths. Also \cite{chen} found the same evidence but they excluded this solution since they assumed that $k_{\perp} \gg k_{\parallel}$ at ion-kinetic scales.
However, other studies show that the $k_{\parallel}$ component of the turbulence, while small compared to $k_{\perp}$, increases around ion-kinetic scales \citep[e.g.][]{bieber1996, leamon1998b, dasso2005, hamilton2008, roberts2015}. In addition, \cite{woodham2018} suggested that this
small $k_{\parallel}$ component is damped from the cascade, which leads to the observed spectral steepening at these scales.
Thus, although observational evidence has been given in favour of the role played by the resonance condition, this problem still deserves a theoretical explanation.

The fact that $k_R$ fits better the kinetic break (at least for the cases studied) does not non exclude however that non-resonant dissipative phenomena could be present, as already stated by \cite{brunotrenchi2014}.

A scaling consistent with the observations has also been found in the numerical hybrid simulations of \cite{franci2016}.
In this work, the authors find that the break length scale $l_b$ can be described at all betas by a combination of $\lambda_i$ and $\lambda_L$ (taking here into account a $2\pi$ difference in the normalisation used in the simulations): $l_b \sim 2 (\lambda_L+ \lambda_i-\sqrt{\lambda_L \lambda_i}/2)$,  leading then to the largest of the two scales at extreme values of $\beta$, as observed by \cite{chen}, and to $l_b > \lambda_L, \lambda_i$ for $\beta\sim1$, consistent with Figure \ref{figure7}.
Although such a scaling is also very similar to $k_R$, it was recovered in 2D simulations with highly oblique k-vectors (main field out-of-plane), so in a condition where cyclotron resonances are not expected to play a significant role.
This might suggest that other processes, related to perpendicular structures, like e.g. magnetic reconnection, could also contribute to the shaping of the spectral transition between MHD and sub-ion scales \citep{franci2017}. 

It is interesting to note that while most of the aspects discussed here (e.g. composition, spectra, microphysics) are well organised by the Alfv\'enicity, suggesting a common solar origin for fast and Alfv\'enic slow streams, this is not the case for the location of the ion-scale spectral break, which on the contrary depends mostly on the plasma beta. This suggests that the evolution of the turbulent cascade at kinetic scales does not depend only on the source properties, but also on the local plasma state and its variation during expansion. As a consequence we find different values of the ion-break in the fast and Alfv\'enic slow regimes, as they are characterized by slightly different beta ranges.

It is worth stressing, within this context, the importance of the role which will be played by future space missions which will investigate the inner heliosphere, i.e. Parker Solar Probe and Solar Orbiter. The former in particular will be extremely useful to study turbulence of this kind of slow wind at an early stage and especially inside the Alfv\'en radius. Actually, following \cite{cranmer2007,cranmer2009} the Alfv\'en radius has a different estimate depending on the solar latitude we are exploring, 20 and 10 solar radii respectively for equatorial regions and poles. 

For a flux tube that passes through a given point
($R_s, \lambda, \phi$) on the source surface, the expansion factor \cite{wang1997} can be calculated. This is the factor by which the flux tube expands in solid angle between its footpoint location ($R_{\odot}, \lambda_0, \phi_0$) and the source surface. The expansion factor f$_s$ equals unity if the bundle of open field lines diverges as r$^2$ but exceeds unity if (as is usually the case) the flux diverges more rapidly than r$^2$. In this case it is called a super radial expansion.

At the borders of coronal holes the Alfv\'en radius is larger than in the fast wind case because of the super-radial expansion. As a matter of fact, Parker Solar Probe will have the possibility to go inside the critical Alfv\'en radius, investigating the Alfv\'enic slow wind, thus giving the opportunity to observe both $z^+$ and $z^-$ and studying the origin of the inward modes. We can take advantage also of possible alignments between Solar Orbiter and Parker Solar Probe in order to analyze the same plasma region across the Alfv\'en radius.

The idea supported by this paper is that this Alfv\'enic slow solar wind would come from low-latitude small coronal holes which are an ubiquitous feature of maximum of solar cycle 23. The results from this study show that the fast wind and the Alfv\'enic slow wind share common characteristics which span from the macrostructure to the microphysics and spectral properties, likely attributable to their similar solar origin i.e. coronal-hole solar wind. It is suggested that in the Alfv\'enic slow wind a major role is played by the super-radial expansion responsible for the lower velocity. In any case, the kind of slow wind which is the main focus of this paper has not only a very high Alfv\'enicity, comparable to that of fast streams ($R_{vb}$ around 0.9 in both cases), but also the amplitude of the fluctuations is very high and comparable to that of fast wind (db/B about 1 in both cases). Remarkably, this seems not to fit well with the typical observations performed at the boundary of coronal holes, when both the Alfv\'enicity and the amplitude of the fluctuations decrease together with the speed \citep[e.g.][]{tsuru2011b}, suggesting then a possible different scenario for the wind regime observed here. Finally, it should be mentioned that some isolated cases of this Alfv\'enic slow wind display a speed profile very similar to that of the fast wind, being characterized by a relatively faster main portion of the stream, followed by a sort of slower rarefaction region. These aspects however need to be investigated further and will benefit from future observations of the inner Heliosphere (e.g. Parker Solar Probe and Solar Orbiter data).

\section*{Acknowledgements}

We acknowledge very useful discussions with Daniele Telloni and Marco Velli.
The authors are grateful to the following people and organizations for data provision: R. Lin (UC Berkeley) and R. P. Lepping (NASA/GSFC) for WIND/3DP and WIND/MFI data, respectively, and G. Gloeckler (University of Maryland) for ACE/SWICS data. All these data are available in the NASA-CDAWeb website: 

\noindent https://cdaweb.sci.gsfc.nasa.gov/index.html.

This research was partially supported by the Agenzia Spaziale Italiana under contracts I/013/12/1 and by the Programme National PNST of CNRS/INSU co-funded by CNES.





\begin{thebibliography}{99}

\bibitem[Abbo et al.(2016)]{abbo} Abbo L., Ofman L., Antiochos S.~K., et al., 2016, \ssr, 201, 55, doi:10.1007/s11214-016-
0264-1, 2016.

\bibitem[Alexandrova et al.(2013)]{alexandrova2013}	Alexandrova O., Bale S.~D. \& Lacombe C., 2013, \prl, 111, 149001

\bibitem[Antiochos et al.(2011)]{antiochos}	Antiochos S.~K., Mikic Z., Titov V.~S. et al., 2011, \apj, 731, 112, doi:10.1088/0004-637X/731/2/112

\bibitem[Antonucci et al.(2005)]{antonucci}	Antonucci E., Abbo L. \& Dodero M.~A., 2005, \aap, 435, 699

\bibitem[Bale et al.(2009)]{bale2009} Bale S.~D., Kasper J.~C., Howes G.~G., et al., 2009, \prl, 103, 211101

\bibitem[Belcher \& Davis(1971)]{bd} Belcher J.~W. \& Davis L., 1971, \jgr, 76, 3534

\bibitem[Belcher \& Solodyna(1975)]{bs} Belcher J.~W. \& Solodyna C.~V., 1975, \jgr, 80, 181

\bibitem[Bieber et al.(1996)]{bieber1996} Bieber J. W., Wanner W., \& Matthaeus W.~H., 1996, \jgr, 101, 2511

\bibitem[Bavassano \& Bruno(1989)]{bavassano1989} 
Bavassano B. \& Bruno R. 1989, \jgr, 94, 977 

\bibitem[Bruno \& Bavassano(1991)]{brubav1991} Bruno R. \& Bavassano B., 1991, \jgr, 96, 7841

\bibitem[Bruno(1992)]{bruno1992} Bruno R., 1992, in Proc. 3rd COSPAR Colloq., Solar Wind Seven, ed. E. Marsch
\& R. Schwenn (Oxford: Pergamon Press), 423

\bibitem[Bruno \& Carbone(2013)]{brunocarbone2013} Bruno R., \& Carbone V., 2013, LRSP, 10, 2

\bibitem[Bruno \& Dobrowolny(1986)]{brunodobro} Bruno R., \& Dobrowolny M., 1986, AnGeo, 4, 17 

\bibitem[Bruno \& Trenchi(2014)]{brunotrenchi2014} Bruno R. \& Trenchi L. 2014, \apj, 787, L24

\bibitem[Bruno et al.(2014)]{bruno2014} Bruno R., Trenchi L., \& Telloni D., 2014, \apj, 793, L15

\bibitem[Bruno et al.(2017)]{bruno2017} Bruno R., Telloni D., De Iure D., Pietropaolo E., 2017, \mnras, 472, 1052

\bibitem[Bruno(2017)]{rb2017} Bruno R., 2017, Parker Solar Probe - Solar Orbiter Joint Meeting, https://sppgway.jhuapl.edu/psp\_SWG\_Oct\_2017\_Mtg

\bibitem[B\"urgi \& Geiss(1986)]{burgi} B\"urgi A., \& Geiss J., 1986, \solphys, 103, 347

\bibitem[Camporeale et al.(2017)]{camporeale2017}
Camporeale, E., Car\'e, A., \& Borovsky, J. E. 2017, JGRA, 122, 10910 

\bibitem[Chen et al.(2014)]{chen} Chen C.~H.~K., Leung L., Boldyrev S., et al., 2014, \grl, 41, 8081

\bibitem[Cranmer(2007)]{cranmer2007} Cranmer S.~R., 2007, in Turbulence and Nonlinear Processes in
Astrophysical Plasmas, Proceedings of the 6th Annual International Astrophysics Conference,
Oahu, Hawaii, 16 – 22 March 2007, (Eds.) Shaikh, D., Zank, G.P., vol. 932 of AIP Conference
Proceedings, pp. 327–332, American Institute of Physics, Melville, NY

\bibitem[Cranmer(2009)]{cranmer2009} Cranmer S.~R., 2009, LRSP, 6, 3

\bibitem[Dasso et al.(2005)]{dasso2005} Dasso S., Milano L.~J., Matthaeus W.~H., \& Smith C.~W., 2005, \apj, 635, L181

\bibitem[D'Amicis et al.(2011)]{damicis2011} D'Amicis R., Bruno R., \& Bavassano B., 2011, JASTP, 73, 653

\bibitem[D'Amicis \& Bruno(2015)]{damicis2015} D'Amicis R. \& Bruno R., 2015, \apj, 805, 84:1

\bibitem[D'Amicis et al.(2016)]{damicis2016} D'Amicis R., Bruno R. \& Matteini L., 2016, in Proceedings of the 14th Solar Wind Conference, AIP Conference Proceedings, edited by L. H. Wang, R. Bruno, E. Moebius, A. Vourlidas, and G. Zank (American Institute of Physics, Melville, NY, 2016) 


\bibitem[Dmitruk \& Matthaeus(2007)]{dmitruk2007} Dmitruk P., \& Matthaeus W.~H., 2007, \pre, 76, 036305

\bibitem[Feldman et al.(1973)]{feldman1973} Feldman W.~C., Asbridge J.~R., Bame S.~J. \&  Montgomery M.~D., 1973, \jgr, 28, 6451

\bibitem[Feldman et al.(1974)]{feldman1974} Feldman, W.~C., Asbridge, J.~R., Bame, S.~J. \&  Montgomery, M.D. 1974, RvGSP, 4, 715

\bibitem[Franci et al.(2016)]{franci2016} Franci L., Landi S., Matteini L. et al., 2016, \apj, 883, 91

\bibitem[Franci et al.(2017)]{franci2017} Franci L., Cerri S.~S., Califano F. et al., 2017, \apjl, 850, L16
 
\bibitem[Fr\"anz et al.(2000)]{franz} Fr\"anz, M., Burgess, D. \& Horbury, T.~S. 2000, JGRA, 105, 12725 

\bibitem[Gary(1993)]{gary1993} Gary, S. P. 1993, in Theory of Space Plasma Microinstabilities (Cambridge: Cambridge Univ. Press), 193

\bibitem[Gary et al.(2002)]{gary2002} Gary, S.~P., Goldstein, B.~E.,\& Neugebauer, M. 2002, JGRA, 107, 4

\bibitem[Geiss et al.(1995)]{geiss1995} Geiss, J., Gloeckler, G., \& von Steiger, R. 1995, SSRv, 72, 49	

\bibitem[Gloeckler et al.(1998)]{gloeckler} Gloeckler, G., Cain, J., \& Ipavich, F. M. 1998, SSRv, 86, 497

\bibitem[Grappin et al.(1991)]{grappin1991} Grappin, R., Velli, M., \& Mangeney, A. 1991, AnGeo, 9, 416

\bibitem[Hamilton et al.(2008)]{hamilton2008}Hamilton, K., Smith, C. W., Vasquez, B. J., \& Leamon, R. J. 2008, JGRA, 113,A01106

\bibitem[He et al.(2015)]{he2015}
He, J., Pei, Z., Wang., L. et al. 2015, ApJ, 805, 176

\bibitem[Hellinger et al.(2006)]{hellinger} Hellinger, P., Tr{\'a}vn{\'{\i}}{\v c}ek, P., Kasper, J.~C. \&
et al. 2006, GeoRL, 33, L09101, doi:10.1029/2006GL025925

\bibitem[Hellinger et al.(2011)]{hellinger2011} Hellinger, P., Matteini, L., {\v S}tver\'ak, S.  et al. 2011, JGRA, 116, 9105

\bibitem[Hellinger et al.(2013)]{hellinger2013} Hellinger, P., Tr{\'a}vn{\'{\i}}{\v c}ek, P., {\v S}tver\'ak, S.  et al. 2013, JGRA, 118, 1351

\bibitem[Hellinger \& Tr{\'a}vn{\'{\i}}{\v c}ek(2014)]{bib8} Hellinger, P. \& Tr{\'a}vn{\'{\i}}{\v c}ek, P. 2014, ApJ, 784, L15

\bibitem[Horbury et al.(2008)]{horbury2008}
Horbury, T.~S., Forman, M., \& Oughton, S. 2008, PRL, 101, 175005

\bibitem[Kasper et al.(2002)]{kasper2002} Kasper, J.~C., Lazarus, A.~J. \& Gary, S.~P. 2002, GeoRL, 29, 1839

\bibitem[Kasper et al.(2008)]{kasper2008} Kasper, J.~C., Lazarus, A.~J. \& Gary, S.~P. 2008, PhRvL, 101, 261103

\bibitem[Ko et al.(2018)]{ko2018} 
Ko, Y.-K., Roberts, D.~A., \& Lepri, S.T. 2018, ApJ, 864, 139

\bibitem[Leamon et al.(1998)]{leamon1998} Leamon, R. J., Matthaeus, W. H., Smith, C. W., \& Wong, H. K. 1998, ApJ, 507, L181

\bibitem[Leamon et al.(1998b)]{leamon1998b} Leamon, R. J., Smith, C. W., Ness, N. F., Matthaeus, W. H., \& Wong, H. K.
1998b, JGRA, 103, 4775

\bibitem[Lepping et al.(1995)]{lepping1995} Lepping, R.~P., Ac{\~u}na, M.~H., Burlaga, L.~F., et al. 1995, SSRv, 71, 207

\bibitem[Lin et al.(1995)]{lin1995} Lin, R.~P., Anderson, K.~A., Ashford, S., et al. 1995, SSRv, 71, 125

\bibitem[Markovskii et al.(2008)]{markovskii2008} Markovskii, S. A., Vasquez, B. J., \& Smith, C. W. 2008, ApJ, 675, 1576

\bibitem[Marsch et al.(1981)]{marsch1981} Marsch, E., Rosenbauer, H., Schwenn, R., et al. 1981, JGR, 86, 9199

\bibitem[Marsch et al.(1982a)]{marsch1982a} Marsch, E., Rosenbauer, H., Schwenn, R., et al. 1982a, JGRA, 87, 35

\bibitem[Marsch et al.(1982b)]{marsch1982b} Marsch, E., Schwenn, R., Rosenbauer, H., et al. 1982b, JGRA, 87, 52

\bibitem[Marsch et al.(1982b)]{marsch1990} Marsch, E., \& Tu, C.-Y. 1990, JGR, 95, 8211

\bibitem[Marsch et al.(2004)]{marsch2004} Marsch, E., Ao, X.-Z. \& Tu, C.-Y. 2004, JGRA, 109, A04102, doi:10.1029/2003JA010330

\bibitem[Marsch (2006)]{marsch2006} Marsch, E. 2006, LRSP, 3, 1

\bibitem[Marsch (2012)]{marsch2012} Marsch, E. 2012, SSRv, 172, 23

\bibitem[Marsch \& Tu(1993)]{marschtu1993}
Marsch, E. \& Tu, C.-Y., 1993, JGR, 98, 21045

\bibitem[Maruca et al.(2012)]{maruca2012} Maruca, B.~A., Kasper, J.~C., \& Gary, S.~P. 2012, ApJ, 748, 137

\bibitem[Matteini et al.(2007)]{matteini2007} Matteini, L., Landi, S. Hellinger, P., et al. 2007, GeoRL, 34, L20105, doi:10.1029/2007GL030920

\bibitem[Matteini et al.(2013)]{matteini2013} Matteini, L., Hellinger, P., Goldstein, B.~E., et al. 2013, JGRA, 118, 2771

\bibitem[Matteini et al.(2014)]{matteini2014} Matteini, L., Horbury, T.~S., Neugebauer, M. \& Goldstein, B.~E. 2014, GeoRL, 41, 259

\bibitem[Matteini et al.(2015)]{matteini2015} Matteini, L. and {Horbury}, T.~S. and {Pantellini}, F. et al. 2015, ApJ, 802, 11

\bibitem[Matteini et al.(2018)]{matteini} 
Matteini, L., Stansby, D., Horbury, T.S. and Chen, C.H.K. 2018, ApJ, submitted

\bibitem[Matthaeus \& Goldstein(1982)]{mg1982} Matthaeus, W.~H., \& Goldstein, M.~L. 1982, JGRA, 87, 6011

\bibitem[Matthaeus \& Goldstein(1986)]{mg1986} Matthaeus, W.~H., \& Goldstein, M.~ L. 1986, PhRvL, 57, 495

\bibitem[Matthaeus et al.(2005)]{matthaeus2005} Matthaeus, W.~H., Dasso, S., Weygand, J.~M., et al. 2005, PhRvL, 95, 1101 

\bibitem[Neugebauer et al.(1996)]{neugebauer1996} Neugebauer, M., Goldstein, B.~E., Smith, E.~J., Feldman, W.~C. 1996, JGRA, 101, 17047

\bibitem[Noci et al.(1997)]{noci} Noci, G., Kohl, J. L., Antonucci, E. et al. 1997, ESA SP-404, 75

\bibitem[Ofman(2000)]{ofman} Ofman, L. 2000, GeoRL, 27, 2885

\bibitem[Platten et al.(2014)]{platten} Platten, S.~J., Parnell, C.~E., Haynes, A.~L., et al. 2014, A\&A, 565, A44

\bibitem[Roberts et al.(1987a)]{roberts} Roberts, D. A., Klein, L.~W., Goldstein, M.~L., \& Matthaeus, W.~H. 1987, JGRA, 92, 11021

\bibitem[Roberts et al.(1987b)]{roberts2} 
Roberts, D. A., Goldstein, M.L., Klein, L.W. \& Matthaeus, W.H. 1987, JGR, 92, 12023 

\bibitem[Roberts \& Li(2015)]{roberts2015}Roberts, O. W., \& Li, X. 2015, ApJ, 802, 1

\bibitem[Sarahoui et al.(2010)]{sarahoui} Sahraoui, F., Goldstein, M.~L., Belmont, G., et al. 2010, PhRvL, 105, 131101

\bibitem[Schwenn(1990)]{schwenn1990} Schwenn, R. 1990, "Large-scale structure of the interplanteray medium" in Physics of the Inner Heliosphere Vol I: Large scale phenomena, Eds Schwenn R ., Marsch E., vol. 20 of Physics and Chemistry in Space, Springer, Berlin; New York

\bibitem[Smith et al.(1978)]{smith1978} 
Smith, E.~J., Tsurutani, B.T., \& Rosenberg, R. L. 1978, JGR, 83, 717 

\bibitem[Smith et al.(2006)]{smith2006} Smith, C.~W., Hamilton, K., Vasquez, B.~J., \& Leamon, R. J. 2006, ApJL, 645, L85

\bibitem[Stakhiv et al.(2015)]{stakhiv2015}
Stakhiv, M., Landi, E., Lepri, S.T., et al. 2015, ApJ, 801, 100

\bibitem[Stansby et al.(2018)]{stansby}
Stansby, D., Horbury, T.S., Matteini, L., MNRAS, accepted

\bibitem[{\v S}tver{\'a}k et al.(2008)]{stverak2008} {\v S}tver{\'a}k, {\v S}., Tr{\'a}vn{\'{\i}}{\v c}ek, P., 
Maksimovic, M., et al. 2008, JGRA, 113, A03103

\bibitem[Telloni et al.(2015)]{telloni2015}
Telloni, D., Bruno, R., \& Trenchi, L. 2015, ApJ, 805, 46

\bibitem[Tsurutani et al.(2002a)]{tsuru2002a}
Tsurutani, B. T., Galvan, C., Arballo, J. K. et al. 2002, GRL, 29, 11, 1528, doi:10.1029GL013623 
 
\bibitem[Tsurutani et al.(2002b)]{tsuru2002b}
Tsurutani, B. T., dasgupta, B., Galvan, C. et al. 2002, GRL, 29, 24, 2233, doi:10.1029/2002GL015652  

\bibitem[Tsurutani et al.(2011a)]{tsuru2011a}
Tsurutani, B. T., Lakhina, G. S., Verkhoglyadova, O. P. et al. 2011, JGRA 116, A02103, doi:10.1029/2010JA015913    

\bibitem[Tsurutani et al.(2011b)]{tsuru2011b}
Tsurutani, B. T., Echer E. \& Gonzalez W. D. 2011, AnnGeo, 29, 839

\bibitem[Tsurutani et al.(2018)]{tsuru2018}
Tsurutani, B. T., Lakhina, G. S., Sen, A., et al. 2018, JGR, 123, 2458, doi.org/10.1002/2017JA024203

\bibitem[Tu \& Marsch(1992)]{tumarsch1992} Tu, C.-Y., \& Marsch, E. 1992, in Proc. 3rd COSPAR Colloq., Solar Wind Seven, ed. E. Marsch
\& R. Schween (Oxford: Pergamon Press), 549

\bibitem[Tu \& Marsch(1994)]{tumarsch1994}
Tu, C.-Y. \& Marsch, E. 1994, JGR, 99, 21481 

\bibitem[Tu \& Marsch(1995)]{tumarsch1995} Tu, C.-Y., \& Marsch, E. 1995, SSRv, 73, 1

\bibitem[Turner et al.(1977)]{turner} Turner, J. M., Burlaga, L.~F., Ness, N.~F., Lemaire, J.~F. 1977, JGRA, 82, 1921

\bibitem[Verdini et al.(2012)]{verdini2012} Verdini, A., Grappin, R., Pinto, R., \& Velli, M. 2012, ApJ, 750, L33

\bibitem[Von Steiger et al.(1997)]{vonsteiger1997} von Steiger, R., J. Geiss, \& G. Gloeckler 1997, in Cosmic Winds and the Heliosphere, edited by J. R. Jokipii, C.~P. Sonett, and M.~S. Giampapa, pp. 581-616, Univ. of Ariz.  Press, Tucson

\bibitem[Von Steiger et al.(2000)]{vonsteiger2000} von Steiger, R., Schwadron, N.~A., Fisk, L.~A., et al. 2000, JGRA, 105, 27217 

\bibitem[Von Steiger(2008)]{vonsteiger2008} von Steiger, R. 2008, in The Heliosphere Through the Solar Activity Cycle, edited by A. Balogh, L.~J. Lanzerotti, and S.~T. Suess, chap. 3, pp. 41–78, doi:10.1007/978-3-540-74302-63, Springer Praxis, Chichester, U. K.

\bibitem[Wang(1994)]{wang1994} Wang, Y.-M. 1994, ApJL, 437, L67

\bibitem[Wang et al.(1997)]{wang1997} Wang, Y.-M., Sheeley, N.~R., Jr., Phillips, J.~L. \& Goldstein, B. E. 1997, ApJL, 488, L51 

\bibitem[Wang \& Shiley(1990)]{wangshiley} Wang, Y.-M., \& Sheeley, N.~R. 1990, ApJ, 365, 372

\bibitem[Wang et al.(2018)]{wang2018}
Wang, X., Tu, C., He, J., \& Wang, L. 2018, ApJ, 857, 136

\bibitem[Winterhalter et al.(1995)]{winterhalter} Winterhalter, D., Neugebauer, M., Goldstein, B.~E. 1995, SSR, 72, 201

\bibitem[Woodham et al.(2018)]{woodham2018} 
Woodham, L.~D., Wicks, R.~T., Verscharen, D., \& Owen, C.~J. 2018, ApJ, 856, 49

\bibitem[Xu \& Borovsky(2015)]{xu2015} 
Xu, F., \& Borovsky, J.~E. 2015, JGR, 120, 70, doi:10.1002/2014JA020412 

\bibitem[Yao et al.(2011)]{yao2011}
Yao, S., He, J.-S., Marsch, E. et al, 2011, ApJ, 728, 146

\bibitem[Zhao et al.(2009)]{zhao2009} 
Zhao, L., Zurbuchen, T.H. \& Fisk, L.A. 2009, GRL, 36, L14104 

\bibitem[Zurbuchen et al.(1999)]{zurbuchen1999} Zurbuchen, T.~H., Hefti, S., Fisk, L.~A., et al. 1999, SSRv, 87,  353

\bibitem[Zurbuchen et al.(2000)]{zurbuchen2000} Zurbuchen, T.~H., Hefti, S., Fisk, L.~A., et al. 2000, JGRA, 105, 18327





\end{thebibliography}








\bsp	
\label{lastpage}
\end{document}